\documentclass[12pt]{iopart}
\pdfoutput=1
\eqnobysec
\usepackage{iopams}
\usepackage{braket}
\usepackage{cite}
\usepackage{iopams}
\usepackage{setstack}
\usepackage[usenames,dvipsnames]{color}
\definecolor{myorange}{RGB}{199.24, 87.48, 47.80}
\usepackage[colorlinks,bookmarks=false,citecolor=NavyBlue,linkcolor=Black,urlcolor=blue]{hyperref}
\usepackage{graphicx}
\newcommand{\be}{\begin{equation}}
\newcommand{\ee}{\end{equation}}
\newcommand{\bea}{\begin{eqnarray}}
\newcommand{\eea}{\end{eqnarray}}
\newcommand{\ba}{\begin{aligned}}
\newcommand{\ea}{\end{aligned}}
\newcommand{\nn}{\nonumber\\}

\usepackage{relsize}
\usepackage{tikz}
\usetikzlibrary{shapes.arrows, fadings}
\tikzfading[name=fade right,
  left color=transparent!0, right color=transparent!100]
  \tikzfading[name=fade left,
  left color=transparent!100, right color=transparent!0]
\definecolor{myorange}{RGB}{199.24, 87.48, 47.80}
\usetikzlibrary{decorations.markings}

\def\doi{http://dx.doi.org/}

\begin{document}
\title{Charges and currents in quantum spin chains:\\
late-time dynamics and spontaneous currents
}
\author{Maurizio Fagotti}
\address{D\'epartement de Physique, \'Ecole Normale Sup\'erieure / PSL Research
University, CNRS, 24 rue Lhomond, 75005 Paris, France}
\ead{mfagotti@phys.ens.fr}
\begin{abstract}
We review  the structure of the conservation laws in noninteracting spin chains and unveil a formal expression for the corresponding currents. 
We briefly discuss how interactions affect the picture.
In the second part, we explore the effects of a localized defect. 
We show that the emergence of spontaneous currents near the defect undermines any description of the late-time dynamics by means of a stationary state in a finite chain. In particular, the diagonal ensemble does not work. 
Finally, we provide numerical evidence that simple generic localized defects are not sufficient to induce thermalization. 
\end{abstract}
\maketitle

\tableofcontents%

\title[Charges and currents in quantum spin chains]{}

\section{Introduction}  %

Local and quasi-local conservation laws and associated currents play a key role in the description of the late-time dynamics after quantum quenches. 
If both the initial state and the Hamiltonian are homogeneous, the stationary properties of local observables can be generally described by the stationary state with maximal entropy under the constraints of the (quasi-)local integrals of motion~\cite{EF:review,P:review}. 
Focusing on integrable models, this picture results in the emergence of so-called generalized Gibbs ensembles (GGE)~\cite{VR:review}. 

On the other hand, if the initial state consists of two semi-infinite homogeneous states joined together, the expectation values of the charges are not sufficient to characterize the late-time dynamics~\cite{BD:review,VM:review}. Nevertheless, in integrable models, the continuity equation satisfied by the charges 
seems to be sufficient to  determine the limit of large time $t$ and large distance $\ell$ from the junction at finite ratio $\zeta=\ell/t$~\cite{BF:16, CAD:hydro,BCNF}. 
In this limit, the expectation values of the local observables become stationary, approaching values dependent only on $\zeta$. The emergent ray-dependent quasi-stationary state was called ``locally-quasi-stationary state'' (LQSS) \cite{BF:16}. 
The continuity equation  puts in relation the LQSS inside the light-cone to the (two, left and right) stationary states that describe expectation values outside. 
There the inhomogeneity of the initial state becomes irrelevant, and the stationary properties are  described by the GGE associated with the homogeneous state on that particular side. The determination of the LQSS is then reduced to the solution of a system of differential equations with given boundary conditions. This method was used in Ref.~\cite{CAD:hydro} and Ref.~\cite{BCNF} to compute the LQSS in the sinh-Gordon quantum field theory and in the XXZ spin-$\frac{1}{2}$ chain, respectively. 

The first part of the paper presents the main characters of the late-time dynamics: conservation laws, currents, and macro-states. 
We review the structure of the conservation laws in noninteracting spin-chain models and calculate the corresponding currents. To the best of our knowledge,  general expressions for the currents have never been reported before.

In the second part, we clarify how charges and currents determine the \mbox{(quasi-)stationary} behavior of local observables and investigate the effects of a localized defect. 
In the presence of a defect, the picture outlined above can be incomplete: some continuity equations develop nontrivial source terms and more boundary conditions are needed, specifically the stationary state in the limit $\zeta\rightarrow 0^\pm$. 
The challenge becomes to understand the dynamics ``close'' to the defect. 
We take a step forward with a proof of impossibility: the stationary behavior of local observables can \emph{not} be described by a genuine stationary state like the \emph{diagonal ensemble}.  
In addition, we provide evidence that simple  integrability-breaking defects do not trigger local thermalization off.  

\subsection{Models}%

The first part of the paper is focussed on noninteracting models. 
Most of the discussion will be general but, for the sake of concreteness, we will consider the explicit example of the quantum XY model in a transverse field
\be\label{eq:XY}
H^{\rm XY}_{\gamma,h}={\mathrm J}\sum_{\ell=-\infty}^\infty \Bigl[\frac{1+\gamma}{4}\sigma_\ell^x\sigma_{\ell+1}^x+ \frac{1-\gamma}{4}\sigma_\ell^y\sigma_{\ell+1}^y+\frac{h}{2}\sigma_\ell^z
\Bigr]\, .
\ee
For $\gamma=1$, this is known as transverse-field Ising chain, while for $\gamma=0$ as XX model. 
Despite being a noninteracting model, the structure of the local conservation laws is particularly various: for generic values of the parameters, $H^{\rm XY}_{\gamma,h}$ has an abelian set of local charges, while for $h=0$ the set becomes non-abelian and there are charges that break one-site shift invariance~\cite{F:super}.  

We will also discuss the Dzyaloshinskii-Moriya (DM) interaction
\be\label{eq:DM}
H_{D}^{\rm DM}=\frac{{\mathrm J} D}{4}\sum_{\ell=-\infty}^\infty(\sigma_\ell^x\sigma_{\ell+1}^y-\sigma_\ell^y\sigma_{\ell+1}^x)\, ,
\ee
which is one of the main causes of the lack of inversion symmetry in spin chains.

As an example of an interacting integrable model, we will consider the spin-$\frac{1}{2}$ XXZ chain, with Hamiltonian
\be\label{eq:HXXZ}
H^{\rm XXZ}_{\Delta}=\frac{{\mathrm J}}{4}\sum_{\ell=-\infty}^\infty\Bigl[\sigma_\ell^x\sigma_{\ell+1}^x+ \sigma_\ell^y\sigma_{\ell+1}^y+\Delta(\sigma_\ell^z\sigma_{\ell+1}^z-1)\Bigr]\, .
\ee
At $\Delta=0$ the model is  noninteracting and corresponds to the XX model. We will use this equivalence to draw a parallel between the conservation laws in interacting integrable and noninteracting spin chains.  

In the final part of the paper the Hamiltonian will be perturbed by defects localized around site $0$.

\subsection{Definitions}  %

Here a list of the definitions that will be used in the text. 

We call ``\emph{support}'' the connected region where a given operator on a spin chain acts nontrivially, \emph{i.e.} differently from the identity. 

We say that $Q$ is ``\emph{localized}'' if its support is finite.

We say that $Q$ is ``\emph{local}'' if  its commutator with any localized operator is localized. 

The ``\emph{range}'' of $Q$ is $1+n$, where $n$ is the maximal increase in the number of sites of the support of a generic localized operator $O$ after taking the commutator with $Q$.

We say that $Q$ is ``\emph{quasi-localized}'' if can be approximated arbitrarily well\footnote{For any $\epsilon> 0$, there is finite $r_\epsilon$ such that the operator distance between $Q$ and a localized operator with range $r_\epsilon$ is smaller than $\epsilon$.} by a localized operator. 

We say that $Q$ is ``\emph{quasi-local}'' if its commutator with any \mbox{quasi-localized} operator is \mbox{quasi-localized}. 

\subsection{Guide to the results}%

The rest of the paper is organized in two almost independent parts, the second part using the results of the first one only in the examples: 
\begin{enumerate}
\item
\begin{itemize}
\item \Sref{s:nonint} is a comprehensive review of the elements of the late-time dynamics in noninteracting spin-$\frac{1}{2}$ chains: charges, currents, and macro-states. 
We supplement the already known structure with a \emph{formal expression for the currents of the local conservation laws}. 
\item \Sref{s:XXZ} draws a parallel between interacting integrable and noninteracting spin chains, using the XXZ spin-$\frac{1}{2}$ chains as archetype of interacting model. 
\end{itemize}
\item
\begin{itemize}
\item \Sref{s:late} is about the late-time dynamics after quantum quenches. 
The section is dedicated to a general formulation of the problem.  
\item \Sref{s:loc} overviews the effects of localized defects in spin chains. It is shown that the stationary behavior of  local observables can not be described by means of stationary states in finite chains, irrespective of how large the chains are. In particular,  \emph{the diagonal ensemble can not be used to describe the late-time expectation values}. In addition, evidence is provided that, also \emph{in the presence of generic localized defects, local observables do not thermalize}. 
\end{itemize}
\end{enumerate}
\begin{itemize}
\item [-]\Sref{s:conc} includes some conclusive remarks and open questions. 
\item [-] \ref{a:nonint} collects further details  on noninteracting spin-chain models. 
\item [-] \ref{a:current} has a proof of the new  formula for the currents.
\end{itemize}
We point out that any section can be skipped without compromising the qualitative comprehension of the next ones, so a reader interested in a particular section is not obliged to follow the route suggested. 

\section{Noninteracting spin-$\frac{1}{2}$ chains}\label{s:nonint} %

This section is dedicated to a particular class of exactly solvable spin-chain models; we aim at giving all the necessary tools to address the problem of non-equilibrium evolution in the simplest many-body quantum systems. 

\subsection{Basics and notations} %

In spin-$\frac{1}{2}$ chains, operators constructed with  the building blocks
\be\label{eq:el}
\sigma_\ell^z\, ,\quad\sigma_\ell^{\alpha_\perp}\Bigl(\prod_{j=\ell+1}^{n-1}\sigma_j^z\Bigr)\sigma_{\ell+n}^{\beta_\perp}\, ,
\ee
with $\alpha_\perp,\beta_\perp\in\{x,y\}$ and $\sigma^j$ being Pauli matrices, are generally said to be ``noninteracting''. This is because the Jordan-Wigner (JW) transformation
\be
\sigma_\ell^x=(-i)^{\ell-1}\prod_{j=1}^{2\ell-1}a_j\, ,\quad \sigma_\ell^y=(-i)^{\ell-1}\prod_{j=1}^{2\ell-2}a_j\, a_{2\ell}\, ,\quad \sigma_\ell^z=-i a_{2\ell-1}a_{2\ell}
\ee
maps the spin operators \eref{eq:el} into \emph{quadratic} forms of Majorana fermions $a_\ell$.  The latter satisfy $\{a_\ell, a_n\}=2\delta_{\ell n}$, where $\{\cdot,\cdot\}$ stands for the anticommutator. We point out that, normally, the Jordan-Wigner transformation is written in terms of the spinless fermions $c^\dag_\ell=(a_{2\ell-1}+i a_{2\ell})/2$; however, the representation in terms of Majorana fermions is more convenient when the number of  fermions $\prod_{j=1}^L(2c^\dag_j c_j-1)$ does not commute with the Hamiltonian (this can be already inferred from the original paper by Lieb, Schultz, and Mattis~\cite{LSM}.)
We focus on spin chains with periodic boundary conditions. 
Crucially, the JW transformation depends explicitly on the choice of the site labelled by $1$.  
This is because noninteracting operators acting around the boundary are not strictly quadratic: \emph{e.g.} $\sigma_L^x\sigma_1^x=i \Pi^z  a_{2L} a_1$, where $L$ is the chain's length and $\Pi^z=\prod_{j=1}^L\sigma_j^z=(-i)^L\prod_{j=1}^{2L}a_j$ measures the parity of the number of fermions.
More generally, a noninteracting operator $Q$ has the form 
\be\label{eq:spinM}
Q=\frac{1-\Pi^z}{2}\frac{1}{4}\sum_{\ell,n=1}^{2L} a_\ell \mathcal Q^+_{\ell n} a_n+\frac{1+\Pi^z}{2}\frac{1}{4}\sum_{\ell,n=1}^{2L} a_\ell \mathcal Q^-_{\ell n} a_n\, ,
\ee 
where $\mathcal Q^\pm$ are purely imaginary Hermitian matrices which can differ only close to the upper-right and lower-left corners. We use capital letters to indicate noninteracting operators, and calligraphic letters to indicate the related matrices. 

For example, the operator $X=\sum_{j=1}^L  \frac{h_j}{2} \sigma_{j}^x\sigma_{j+1}^z\sigma_{j+2}^x$, with $\sigma^\alpha_{L+j}\equiv \sigma^\alpha_j$, is noninteracting, and can be written as follows
\be\label{eq:example}
X=-i \sum_{j=1}^{L-2}  \frac{h_j}{2}  a_{2j} a_{2j+3}+i\frac{h_{L-1}}{2} \Pi^z a_{2L-2} a_{1}^x+i \frac{h_L}{2} \Pi^z a_{2L}a_{3}^x\, .
\ee
The corresponding matrices $\mathcal  X^\pm$ are given by
\bea
 {\mathcal X}^\pm_{\ell n}=&-i\frac{1+(-1)^\ell}{2}h_{\ell/2} \delta_{n, \ell+3}+i\frac{1-(-1)^\ell}{2}h_{(\ell-3)/2} \delta_{\ell, n+3}\nn
&\mp  i  h_{L-1}(\delta_{\ell, 2L-2} \delta_{n 1}-\delta_{n, 2L-2} \delta_{\ell 1})\mp i h_{L}(\delta_{\ell, 2L} \delta_{n 3}-\delta_{n, 2L} \delta_{\ell 3})\, .
\eea

If $Q$ is translationally invariant (in the previous example, $h_j=h$), $\mathcal Q^\pm$ are
\mbox{block-(anti-)circulant} matrices:
\begin{eqnarray}
\mathcal Q^\pm _{\ell+2\kappa, n+2\kappa}=\mathcal Q^\pm _{\ell n}\qquad 1\leq \ell,n\leq 2L-2\kappa\\
\mathcal Q^\pm_{\ell+2L, n}=\pm \mathcal Q^\pm_{\ell, n}\, .
\end{eqnarray}
Here $\kappa$ is the number of sites of the elementary shift under which the spin operator is invariant. 
We note that $\Pi^z$ commutes with the noninteracting operators \eref{eq:el} and, in particular, with $Q$. 

The structure of $\mathcal Q^\pm$ becomes manifest in the Fourier space:
\be\label{eq:symb}
\fl\quad \mathcal Q^{\pm}_{2\kappa \ell+i, 2\kappa n+j}=\frac{\kappa}{L}\sum_{p|e^{i p L/\kappa}=\pm 1}e^{i(\ell-n) p}[\hat q_\kappa(e^{i p})]_{i j}\qquad 0\leq \ell,n\leq \frac{L}{\kappa}\, ,\quad 1\leq i,j\leq 2\kappa\, .
\ee
Here we introduced the $\kappa$-site \emph{symbol} $\hat q_\kappa(z)$, which is a $(2\kappa)$-by-$(2\kappa)$ matrix encoding any information about $Q$. 
For the symbol of a noninteracting operator, we use the corresponding  lower-case letter, with an additional hat. 
The Hermiticity of the operator plus the algebra of the Majorana fermions $\{a_\ell,a_n\}=2\delta_{\ell n}$ implies
\be\label{eq:symbprop}
\hat q_\kappa(z)=-\hat q_\kappa^\ast(z)=-\hat q_\kappa^t(1/z)\, .
\ee 
We indicate by $\hat q^{n}_\kappa$ the coefficients of the inverse Fourier transform:
\be
\hat q_\kappa(z)=\sum_n z^n \hat q^{n}_\kappa\, ,
\ee
\be
[\hat q^{n}_\kappa]_{i j}=\left\{\begin{array}{ll}
\mathcal Q^{\pm}_{i, 2\kappa n+j}&n\geq 0\\
\mathcal Q^{\pm}_{-2\kappa n+i,j}&n<0\\
\end{array}\right.\qquad 1\leq i,j\leq 2\kappa\, .
\ee
In the example \eref{eq:example} with $h_j=h$, the one-site symbol is $\hat {x}_1(e^{i p})=-h\sigma^y e^{2i p\sigma^z}$, while the two-site symbol reads as $\hat{x}_2(e^{i p})=-h \mathrm I_2\otimes [\sigma^y e^{i p\sigma^z}]$, where $\mathrm I_2$ is the $2$-by-$2$ identity. 
If the symbol is linear in $z$ and $1/z$ (\emph{i.e.} $\mathcal Q^{\pm}$ are block-tridiagonal with $(2\kappa)$-by-$(2\kappa)$ blocks) we will also write $\hat q_\kappa^-$ instead of $\hat q_\kappa^{-1}$ and $\hat q_\kappa^+$ instead of $\hat q_\kappa^{1}$.

The spectrum of $Q$ can be split in two sectors, depending on the eigenvalue of $\Pi^z$. Each sector is generated by the single particle dispersion relations, which consist of the positive eigenvalues of $\mathcal Q^\pm$ and are equal to the positive eigenvalues of the symbol $\hat q_\kappa(e^{i p})$ computed in the allowed momenta $e^{ip L/\kappa}=\pm 1$. 
We note that two excited states in the same sector differ only in an even number of elementary excitations; an odd number of excitations would change the eigenvalue of~$\Pi^z$.

For future convenience, we report the one-site symbol of the XY Hamiltonian \eref{eq:XY} with the DM interaction \eref{eq:DM}, $H=H^{\rm XY}_{\gamma,h}+H^{\rm DM}_{D}$:
\be\label{eq:XYsymb}
\hat h_{1}^{(\gamma,h,D)}(e^{i p})={\mathrm J}\bigl[(h-\cos p)\sigma^y+\gamma\sin p\, \sigma^x+D\sin p\, \mathrm I\bigr]\, .
\ee
Thus we have
\be
[\hat h_{1}^{(\gamma,h,D)}]^{\pm}=-\frac{{\mathrm J}}{2}(\sigma^y\pm i \gamma\sigma^x\pm i D\mathrm I)\, ,\qquad
[\hat h_{1}^{(\gamma,h,D)}]^{0}={\mathrm J} h\sigma^y\, .
\ee

\subsubsection{(Quasi-)Locality.}  %

Translationally invariant noninteracting operators $Q$ are local if $\hat q_\kappa(z)$ is a (matrix) polynomial in $z$ and $1/z$. This follows directly from the definition \eref{eq:symb}, indeed the support of $a_{2\ell+i} a_{2n+j}$, with $i,j\in\{1,2\}$,  consists of the sites between $\ell$ and $n$. Thus, the range of the operator is simply given by one plus $\kappa$ times the degree of the polynomial. 

More generally, if $\hat q_\kappa(e^{i p})$ is a smooth $2\pi$-periodic function of $p$, $Q$ is \emph{quasi-local}, having tails that decay faster than any power of the distance.

\subsubsection{Expectation values.}%

For the sake of simplicity, let us consider the system in the thermodynamic limit $L\rightarrow\infty$. 

We focus on a state of the form 
\be\label{eq:Gaussian}
\rho=\frac{e^{Q}}{\mathrm{tr}[e^{Q}]}\, ,
\ee
with $Q$ a noninteracting operator as in \eref{eq:spinM}. 
This could be a thermal state ($Q=-\beta H$) but also the ground state of a noninteracting Hamiltonian if the spin-flip symmetry generated by $\Pi^z$ is not broken (then, the ground state can be obtained as the low-temperature limit of thermal states). 

The expectation value of any operator consisting of an odd number of fermions vanishes. On the other hand, by Wick's theorem, the expectation value of an even number of fermions can be reduced as follows
\be
\mathrm{tr}[\rho a_{i_1}\cdots a_{i_N}]=\mathrm{pf}[A_{\{i_1,\dots,i_N\}}]\qquad i_j=i_{j'}\Leftrightarrow j=j'\, .
\ee
Here $\mathrm{pf}$ denotes the Pfaffian and $A_{\{i_1,\dots,i_N\}}$ is the antisymmetric matrix with upper triangular elements given by
\be
{}[A_{\{i_1,\dots,i_N\}}]_{\ell, n}=\mathrm{tr}[\rho a_{i_\ell} a_{i_n}]\qquad 1\leq \ell<n\leq N\, .
\ee
The matrix
\be
\Gamma_{i j}=\mathrm{tr}[\rho a_i a_j]-\delta_{i j}
\ee
is the (fermionic) correlation matrix and is also known as propagator or Green's function. If $Q$ is translationally invariant,  $\Gamma$ is a block-Toeplitz matrix with symbol~\cite{FC:disjoint}
\be\label{eq:Gamma}
\hat \Gamma_\kappa(z)=\tanh\Bigl(\frac{\hat q_\kappa(z)}{2}\Bigr)\, ,
\ee
where $\hat q_\kappa(z)$ is the symbol of $Q$.
The expectation value of a noninteracting \mbox{(quasi-)localized} operator $O_\ell$  is then given by
\be\label{eq:EV}
\mathrm{tr}[\rho O_\ell]=\frac{1}{4\kappa}\int_{-\pi}^\pi\frac{\mathrm d p}{2\pi} \mathrm{tr}[\hat\Gamma_\kappa(e^{i p}) \hat o_\kappa(e^{i p})]\, ,
\ee
where $\hat o_\kappa(z)$ is the symbol of the translationally invariant (quasi-)local operator $O=\sum_\ell O_\ell$.   

\subsection{The elements of the dynamics}  %

We discuss here the quantities that play a key role in the dynamics.  

\subsubsection{Local conservation laws.}\label{s:charges} %

The (noninteracting) conservation laws have the form~\eref{eq:spinM}, and have symbols $\hat q_\kappa(z)$ that commute with the symbol $\hat h_\kappa(z)$ of the Hamiltonian~\cite{RDM,F:super}:
\be\label{eq:comm}
[\hat q_\kappa(z),\hat h_\kappa(z)]=0\, .
\ee
 Thus, the set of the local conservation laws is obtained by identifying the most general symbols which satisfy \eref{eq:comm} and are polynomials in $z$ and $1/z$. 
 This is generally a very simple task and the interested reader can find more details in \ref{a:nonint}.
 We just note that there is a simple way to generate charges with increasing range: if $\hat q_\kappa(z)$ is the symbol of a local charge, $\frac{z^n+z^{-n}}{2} \hat q_\kappa(z)$ is as well, with a support extending over $\kappa n$ further sites. 
 
In the specific case of the XY model with $h\neq 0$, the local conservation laws are organized in two classes~\cite{Ischarge}: $I^{(n,+)}$ and $I^{(n,-)}$, with $n\geq 0$ being the range minus two. The corresponding symbols are given by~\cite{RDM}
\be\label{eq:chargesXY}
\hat\imath^{(n,+)}_{1}(e^{i p})=\cos(n p)\hat h_1^{(\gamma,h,0)}(e^{i p})\, ,\qquad \hat\imath^{(n,-)}_{1}(e^{i p})={\mathrm J} \sin((n+1) p)\mathrm I_2\, .
\ee
In addition, in the isotropic case ($\gamma=0$), the total spin in the $z$ direction is conserved; its one-site symbol is $\hat\imath^{(0,+)}_{1}(e^{i p})/\cos p$. 

We note that the DM interaction~\eref{eq:DM} commutes with the XY Hamiltonian; its  one-site symbol  is 
$D\, \imath^{(0,-)}_{1}(e^{i p})$. 

As shown in \ref{a:nonint}, for $h=0$ one can find additional two-site shift invariant charges~\cite{F:super}, which manifest the existence of two-site symbols, polynomial in $z$ and $1/z$,  which commute with $\hat h_2(z)$ but do not commute with one another.
Specifically, besides $I^{(n,+)}$ and $I^{(n,-)}$, Ref.~\cite{F:super} found new local charges $Y^{(n,s_1,s_2)}$, which are odd under a one-site shift and take sign $s_1$ under chain inversion $\rm R$ and sign $s_2$ under spin flip $\Pi^x: O\mapsto \prod_\ell\sigma_\ell^x O \prod_\ell\sigma_\ell^x$. Their two-site symbols (not being one-site shift invariant, they do not have a one-site symbol) read as
\begin{eqnarray}\label{eq:QXY}
\fl\qquad\qquad \hat y_2^{(n,++)}(e^{i p})=\cos(n p)[\sigma^ye^{i\frac{p}{2}\sigma^z}]\otimes [i\sigma^z\hat h_{1}^{(\gamma,0,0)}(e^{i p/2})]\nn
\fl\qquad\qquad \hat y_2^{(n,+-)}(e^{i p})={\mathrm J}\cos((n+1/2) p)[\sigma^ye^{i\frac{p}{2}\sigma^z}]\otimes \sigma^z\nn
\fl\qquad\qquad \hat y_2^{(n,-+)}(e^{i p})={\mathrm J}\sin((n+1) p)\sigma^z \otimes \sigma^z\nn
\fl\qquad\qquad \hat y_2^{(n,--)}(e^{i p})=\sin((n+1/2) p)\sigma^z  \otimes [i\sigma^z\hat h_{1}^{(\gamma,0,0)}(e^{i p/2})]\, .
\end{eqnarray}
Remarkably, in \ref{a:nonint} it is shown that the full set of local charges $\{I^{(n,s)},Y^{(n,s_1,s_2)}\}$ can be reorganized in quasi-local operators $T^{(n,\alpha)}$ which generate the loop algebra
\be\label{eq:loop}
{}[T^{(m,\alpha)},T^{(n,\beta)} ]=2i \sum_\gamma\epsilon_{\alpha\beta\gamma}  T^{(m+n,\gamma)}\, ,
\ee
where
\be
\epsilon_{\alpha\beta\gamma}=(1-\delta_{\alpha 0})(1-\delta_{\beta 0})(1-\delta_{\gamma 0})\varepsilon_{\alpha\beta\gamma}
\ee
and $\varepsilon_{\alpha\beta\gamma}$ is the Levi-Civita symbol. 
While the existence of an $sl_2$ loop-algebra symmetry was known long before~\cite{DFM:2000}, the existence of quasi-local generators in the quantum XY model has been realized only recently~\cite{F:super, Bruno}. 
 
 \subsubsection{Currents.}\label{s:current}%

 In a spin chain, the current $J[Q]$ of a charge $Q$ satisfies 
 \be\label{eq:continuity}
J_{\ell+n}[Q]-J_{\ell}[Q]=-i\Bigl[H,\sum_{j=\ell}^{\ell+n-1} Q_j\Bigr]\, ,
 \ee
 where $Q_\ell$ and $J_{\ell}[Q]$ are the charge density and the current density, respectively. 
We note that the current is defined up to a constant, which we choose in such a way to make the operator traceless.
 
 In \ref{a:current} it is shown that the (noninteracting) current $J[Q]$ of a (noninteracting) local conservation law $Q$ has the symbol 
\be\label{eq:current}
\hat \jmath_\kappa(z)= \frac{1}{2}\{\hat q_\kappa(z),i z\partial_z\hat h_\kappa(z)\}\quad \mathrm{i.e.}\quad\hat \jmath_\kappa(e^{i p})= \frac{1}{2}\{\hat q_\kappa(e^{i p}),\partial_p\hat h_\kappa(e^{i p})\}\, ,
\ee
where $\{\cdot,\cdot\}$ stands for the anticommutator. 
To the best of our knowledge, this has never been pointed out before.  From \eref{eq:current}, one can immediately obtain the current associated with a generic noninteracting conservation law, which in the thermodynamic limit reads as 
\be\label{eq:spinJ}
\fl\qquad\qquad J[Q]=\frac{1}{8} \sum_{\ell,n=-\infty}^\infty\sum_{i,j=1}^{2\kappa}  \int\frac{\mathrm d p}{2\pi}e^{i(\ell-n) p}\, \{\hat q_\kappa(e^{i p}),\partial_p\hat h_\kappa(e^{i p})\}_{i j}\  a_{2\kappa \ell+i} a_{2\kappa n+j}\, .
\ee 
The expression in terms of spins can be obtained using the inverse JW transformation
\be
a_{2\ell-1}=\prod_{j=1}^{\ell-1}\sigma_j^z\, \sigma_\ell^x\, ,\qquad a_{2\ell}=\prod_{j=1}^{\ell-1}\sigma_j^z\, \sigma_{\ell}^y\, .
\ee
The importance of having an exact expression for a generic current will become clear in \sref{s:late}, where the relation between charges and currents will be used to determine the (quasi-)stationary behavior of local observables. 

\paragraph{Example.}%

We consider the specific case of the XY Hamiltonian with the DM interaction, which has the symbol~\eref{eq:XYsymb}. For generic values of the parameters the symbols of the local conservation laws are given by \eref{eq:chargesXY}. 
Using \eref{eq:current}, the symbols of the currents of the reflection symmetric charges $I^{(n,+)}$ are given by
\be\label{eq:Jn+}
\hat \jmath^{(n,+)}_{1}(e^{i p})=
\cos(np)(v_{(\gamma,h,0)}(p)\varepsilon_{(\gamma,h,0)}(p)\mathrm I_2+{\mathrm J} D\cos p\, \hat h_1^{(\gamma,h,0)}(e^{i p}))\, ,
\ee
where $\varepsilon_{(\gamma,h,0)}(p)={\mathrm J}\sqrt{(h-\cos p)^2+\gamma^2\sin^2 p}$ is the dispersion relation of the XY model and $v_{(\gamma,h,0)}(p)=\varepsilon_{(\gamma,h,0)}'(p)$ is the excitation velocity.
Remarkably, $\hat \jmath^{(n,+)}_1(z)$ commutes with the symbol of the Hamiltonian, that is to say the currents corresponding to the charges $I^{(n,+)}$ are conserved. 
Explicitly, they are given by
\begin{eqnarray}
J[I^{(n,+)}]=\frac{\mathrm J}{2}\Bigl[&h(I^{(n,-)}-I^{(n-2,-)})+\frac{\gamma^2-1}{2}(I^{(n+1,-)}-I^{(n-3,-)})+\nn&D(I^{(n+1,+)}+I^{(n-1,+)})\Bigr]\, .
\end{eqnarray}
The symbols of the currents associated with the charges  $I^{(n,-)}$, odd under reflection symmetry, are instead
\be\label{eq:Jn-}
\hat \jmath^{(n,-)}_1(e^{i p})=
{\mathrm J} \sin((n+1) p)({\mathrm J} D \cos p\, \mathrm I_2+\hat h_1^{(\gamma,0,0)}(ie^{i p}))\, .
\ee
The corresponding currents are \emph{not} conserved. Explicitly, they are given by
\bea
\fl J[I^{(n,-)}]= \frac{{\mathrm J}^2}{2} \Bigl\{D(I^{(n+1,-)}+I^{(n-1,-)})-\nn
\fl \quad i\sum_\ell\Bigl[ \frac{1+\gamma}{4}(a_{2\ell}a_{2\ell+2n+3}+a_{2\ell-1}a_{2\ell+2n})-\frac{1-\gamma}{4}(a_{2\ell}a_{2\ell+2n-1}+a_{2\ell-1}a_{2\ell+2n+4})\Bigr]\Bigr\}\, .
\eea

\subsubsection{Macro-state.}\label{ss:macro} %

In the thermodynamic limit, infinitely many excited states share the same thermodynamic properties. They are locally indistinguishable and are said to form a macro-state~\cite{EF:review}. In integrable models, the macro-state is characterized by a set of densities $\{\rho_{n,{\rm p}}(\lambda)|n=1,2,\dots\}$, where $n$ labels all distinct species of stable excitations in the model. 
We will indicate a generic state which represents the macro-state by $\ket{\rho}$. 
This can be a pure state~\cite{C:review}, but also a generalized Gibbs ensemble $\frac{1}{Z}e^{-H_{\rm GGE}}$, with $H_{\rm GGE}$ a linear combination of the relevant conservation laws $[H_{\rm GGE},H]=0$ - \sref{ss:GGE}. 
If the set of the relevant charges is abelian, $\ket{\rho}$ is by construction a simultaneous eigenstate of such conservation laws, and the eigenvalue per unit length of a charge $Q$ can be generally written as the sum of the contributions of each excitation, weighted with the corresponding density:
\be\label{eq:root}
\braket{\rho|Q_\ell|\rho}=\sum_n \int \mathrm d \lambda q_n(\lambda)\rho_{n,{\rm p}}(\lambda) +\braket{0|Q_\ell|0}\, .
\ee
Here $q_n(\lambda)$ are functions independent of $\rho_{n,{\rm p}}$ and $\ket{0}$ is called the reference state. 

\paragraph{Example.}%

We now focus on the XY model for generic values of the parameters, but almost any result can be easily  generalized to other noninteracting models. 
By \eref{eq:Gamma}, the symbol of the correlation matrix is a function of the Hamiltonian symbol, and hence is written as a linear combination of the symbols of the local conservation laws. In addition, it has eigenvalues between $-1$ and $1$.
The symbol of the correlation matrix of a one-site shift invariant macro-state can then be parametrized as follows
\be\label{eq:Gammarho}
\fl\qquad\qquad \hat \Gamma_1(e^{i p})=\frac{2\pi\rho_{1,{\rm p}}(p)+2\pi\rho_{1,{\rm p}}(-p)-1}{\varepsilon_{(\gamma,h,0)}(p)}\hat h_1^{(\gamma,h,0)}(e^{i p})+2\pi(\rho_{1,{\rm p}}(p)-\rho_{1,{\rm p}}(-p)) {\rm I}\, ,
\ee
where $0\leq \rho_{1,{\rm p}}(k)\leq \frac{1}{2\pi}$ is $2\pi$-periodic.

The integrals of motion have the form \eref{eq:root}, indeed, using \eref{eq:EV}, we find
\begin{eqnarray}
\braket{\rho|I_\ell^{(n,+)}|\rho}&=\int_{-\pi}^\pi\mathrm d p\cos(n p)\varepsilon_{(\gamma,h,0)}(p)\Bigl[\rho_{1,{\rm p}}(p)-\frac{1}{4\pi}\Bigr]\\
\braket{\rho|I_\ell^{(n,-)}|\rho}&={\mathrm J}\int_{-\pi}^\pi \mathrm d p\sin((n+1)p)\rho_{1,{\rm p}}(p)\, .
\end{eqnarray}
The energy density $I_\ell^{(0,+)}$ is minimized by $\rho_{1,{\rm p}}(p)=0$, therefore the reference state of the parametrization proposed is the ground state of the XY Hamiltonian. 
En passant, we note that this description in terms of a single species of excitations is just a matter of conventions and it is not unusual to describe the macro-state introducing more densities (in particular, the XXZ model \eref{eq:HXXZ} with $\Delta=0$ is usually characterized by two root densities.) 
The functions associated with the charges appearing in \eref{eq:root} are  given by 
\be
q_1^{(n,+)}(p)=\cos(n p)\varepsilon_{\rm XY}(p)\qquad q_1^{(n,-)}(p)={\mathrm J}\sin((n+1) p)\, .
\ee
Using \eref{eq:Jn+} and \eref{eq:Jn-}, we can easily compute the expectation values of the currents for the XY model with the DM interaction. We find
\begin{eqnarray}
 \braket{\rho|J_{\ell}^{(n,+)}|\rho}&=\int_{-\pi}^\pi\mathrm d p\,  v_{(\gamma,h,D)}(p) \cos(n p)\varepsilon_{(\gamma,h,0)}(p)\rho_{1,{\rm p}}(p)+\mathrm{const.}\\
 \braket{\rho|J_{\ell}^{(n,-)}|\rho}&=\int_{-\pi}^\pi\mathrm d p\, v_{(\gamma,h,D)}(p) {\mathrm J} \sin((n+1) p)\rho_{1,\rm{p}}(p)+\mathrm{const.}
\end{eqnarray}
where $v_{(\gamma,h,D)}(p)=v_{(\gamma,h,0)}(p)+{\mathrm J} D \cos p$ is the velocity of the quasiparticles for the XY+DM Hamiltonian.

From these results we deduce that, for a generic charge $Q$, the current is simply obtained multiplying the integrand by the velocity $v_1(k)$ of the quasi-particle excitations:
\begin{eqnarray}
\braket{\rho|Q_\ell|\rho}&=\int_{-\pi}^\pi\mathrm d p\, q_1(p)\rho_{1,{\rm p}}(p)+\mathrm{const.}\nn
\braket{\rho|J_{\ell}[Q]|\rho}&=\int_{-\pi}^\pi\mathrm d p\, q_1(p) v_1(p) \rho_{1,\rm{p}}(p)+\mathrm{const.}\label{eq:rhocurr}
\end{eqnarray}
We stress that the expectation values of the currents take this simple form only because $\ket{\rho}$ is stationary.
These expressions agree with the expectations based on semiclassical arguments~\cite{semi-cl,Viti_inhom,BF:16, BCNF}, where the time variation of a charge density can be interpreted as the result of quasi-particle excitations moving throughout the chain. 

Finally, we point out that  the parametrization in terms of  densities can be more complicated when the set of charges is non-abelian. Some details are reported in \ref{a:nonab}.

\subsection{Dynamics.} %

Time evolution under a noninteracting Hamiltonian $H$ preserves the noninteracting structure: if $O(t)=e^{i H t}O e^{-i H t}$ is a noninteracting operator in the Heisenberg picture, we have (\emph{cf}. \eref{eq:spinM})
\be
\fl\qquad\qquad i \partial_t O(t)=[O(t),H]=\frac{1-\Pi^z}{8}\vec a^\dag [\mathcal O^+(t),\mathcal H^+]\vec a+\frac{1+\Pi^z}{8}\vec a^\dag  [\mathcal O^-(t),\mathcal H^-]\vec a\, ,
\ee
where we introduced the vector notations $[\vec a]_\ell =a_\ell$. 
Thus, also the matrices associated with the operators satisfy 
\be
i\partial_t{\mathcal O}^\pm(t)=[\mathcal O^\pm(t),\mathcal H^\pm]\, .
\ee
If $O$ is translationally invariant, this can be reduced to an equation for the symbol
\be
i\partial_t\hat o_\kappa(z,t)=[\hat o_\kappa(z,t),\hat h_\kappa(z)]
\ee
\emph{i.e.}
\be
\hat o_\kappa(z,t)=e^{i \hat h_\kappa(z) t}\hat o_\kappa(z)e^{-i \hat h_\kappa(z) t}\, ,
\ee
where we wrote $\hat o_\kappa(z)$ instead of $\hat o_\kappa(z,0)$. 
Using \eref{eq:EV}, in the thermodynamic limit, the time evolution of the expectation value of a noninteracting (quasi-)local operator $O_\ell$, in a translationally invariant state where the symbol of the correlation matrix is given by $\hat\Gamma_\kappa(z)$, reads as
\be\label{eq:Ot}
\mathrm{tr}[e^{- i H t}\rho e^{i H t}O_\ell]=\frac{1}{4\kappa}\int_{-\pi}^\pi\frac{\mathrm d p}{2\pi} \mathrm{tr}[e^{-i \hat h_\kappa(e^{ip}) t}\hat\Gamma_\kappa(e^{i p}) e^{i \hat h_\kappa(e^{ip}) t}\hat o_\kappa(e^{i p})]\, .
\ee
Being $O_\ell$ arbitrary, we also see that, in the Schr\"odinger picture, the symbol of the correlation matrix time evolves as follows
\be
\Gamma_\kappa(e^{i p},t)=e^{-i \hat h_\kappa(e^{ip}) t}\hat\Gamma_\kappa(e^{i p}) e^{i \hat h_\kappa(e^{ip}) t}\, .
\ee 
If the dispersion relation does not have flat parts, by the Riemann-Lebesgue lemma, in \eref{eq:Ot} only the time-independent contributions survive the limit of infinite time, and  $\hat\Gamma_\kappa(z,t)$ can be replaced by a linear combination of the symbols of the charges, like in \eref{eq:Gammarho}. 
That is to say, \emph{the late-time dynamics can be described by a macro-state}. 
This is the gist of Ref.~\cite{TFIC0}. 

This simple but powerful result relies on translational invariance.
However, the emergence of (quasi-)stationary behavior is not restricted to homogeneous systems, as will be discussed in \sref{ss:LQSS}.

\section{Adding interactions}\label{s:XXZ} %

Before investigating the effects of inhomogeneities and defects on the dynamics after quantum quenches, it is fundamental to clarify how interactions affect the results of the previous section. In fact, while the framework presented can only be applied to noninteracting models, most of the conclusions hold true also in the presence of interactions preserving integrability.  

In the next subsections we draw a concise parallel between interacting integrable and noninteracting spin chains, using the XXZ spin-$\frac{1}{2}$ chain as archetype of interacting model. 

\subsection{Conservation laws} \label{ss:chargeXXZ}  %

An important class of exactly solvable models is solvable by the algebraic Bethe ansatz method~\cite{Korepinbook}.  Local conservation laws $Q^{(n)}$ are obtained from the derivatives of the logarithm of a transfer matrix $\tau (\lambda)$, computed at a particular point $\lambda_0$, known as shift point. The name is because  $\tau (\lambda_0)$ shifts the operators on the chain by one site.
The transfer matrix commutes at different values of the spectral parameter $\lambda$, \emph{i.e.} $[\tau(\lambda),\tau(\lambda')]=0$, so $Q^{(n)}$ form an abelian set of local conservation laws $[Q^{(n)},Q^{(m)}]=0$. 
 
In particular, for the XXZ model~\eref{eq:HXXZ} one has
\be\label{eq:locXXZ}
Q_n=i\Bigl(\frac{\sin \gamma}{\gamma}\frac{\partial}{\partial \lambda}\Bigr)^{n-1}\log\tau(i+\lambda)\Bigr|_{\lambda=0}
\ee
where $\gamma=\arccos \Delta$ and
\begin{eqnarray}\label{eq:tm}
\tau(i+\lambda)=\mathrm{tr}_{0}\Bigl[\prod_{j=L}^1\mathcal L_j(\lambda)\Bigr]\\
\fl  \mathcal L_j(\lambda)=\frac{\mathrm I_j \mathrm I_0+\sigma^z_j\sigma^z_0}{2}+\frac{\sin(\frac{i\gamma\lambda}{2})}{\sin(\frac{i\gamma\lambda}{2}-\gamma)}\frac{\mathrm I_j \mathrm I_0-\sigma^z_j\sigma^z_0}{2}-\frac{\sin(\gamma)}{\sin(\frac{i\gamma\lambda}{2}-\gamma)}(\sigma^+_j\sigma_0^-+\sigma^-_j\sigma_0^+)\, .
\end{eqnarray}
Here we labeled the physical sites with integers from $1$ to $L$, while  site zero is the auxiliary space over which the trace is taken.  
The Hamiltonian is $H^{\rm XXZ}_{\Delta}=Q_1$.

The charges~\eref{eq:locXXZ} are one-site shift invariant and, under reflections,  transform as $Q_n\rightarrow -(-1)^n Q_n$. 
In addition, they are invariant under spin flip $\Pi^x$, indeed the latter can be reduced to a unitary transformation in the auxiliary space 
\be
\Pi^x:\quad \sigma_0^\alpha \rightarrow \sigma_0^x \sigma_0^\alpha \sigma_0^x\, .
\ee
Since the Hamiltonian is invariant under rotations about the $z$ axis, it also commutes with the total spin in the $z$ direction $S^z=\frac{1}{2}\sum_\ell\sigma_\ell^z$, which is independent of the other charges $\{Q_n\}$.
For long time these have been the only charges known with nice local properties. However, setting $\Delta=0$ is sufficient to infer that the set is  not generally complete. Indeed, for $\Delta=0$ the XXZ model is reduced to the XX model (XY model with $\gamma=h=0$), where we found~(\emph{cf.} \eref{eq:QXY})
\begin{itemize}
\item half of the families of local charges are odd under spin flip:

$\{I^{(2n+1,+)},I^{(2n,-)},Y^{(n,+-)},Y^{(n,--)}\}$
\item half of the families of local charges are odd under a shift by one site:

$\{Y^{(n,++)},Y^{(n,+-)},Y^{(n,-+)},Y^{(n,--)}\}$
\end{itemize}
Are these properties peculiar to the noninteracting limit? 

\paragraph{Odd charges under spin flip.}
The existence of charges that are odd under spin flip in the gapless phase $|\Delta|<1$ was suspected for long times, since the spin Drude weight is positive, despite the fact that, at zero magnetization, the spin current is orthogonal to the $Q_n$ \cite{H-M,Karrash1,Karrash2}. It has been then shown that, at the so-called root of unity values $\gamma=\pi r$, with $r$ a rational number, it is possible to construct families of charges which are odd under spin flip~\cite{ql_prosen,ql_pereira}. The contruction is based on the fact that the transfer matrix \eref{eq:tm} is part of a much larger family of commuting operators with
\be
\mathcal L_j(z)=\frac{1}{2}\left(\begin{array}{cc}
z K-z^{-1}K^{-1}&2i z\sin\gamma S^-\\
2i z^{-1}\sin\gamma S^+&z K^{-1}-z^{-1}K
\end{array}\right)\, ,
\ee
where the matrix represents the (quantum) space of the $j$-th spin, while  
the matrix elements act on the auxiliary space. The operators $K$, $S^+$ and $S^-$ satisfy the quantum group algebra $U_{q}[SU(2)]$
\begin{eqnarray}\label{eq:qga}
K S^+=e^{i\gamma}S^+ K\nn
K S^-=e^{-i\gamma}S^- K\nn
{}[S^+,S^-]=\frac{K^2-K^{-2}}{2i\sin\gamma}\, .
\end{eqnarray}
It was shown that non-unitary representations of the quantum group generate transfer matrices which are not invariant under spin-flip. Specifically, one can choose
\be
\fl K\ket{j}=e^{i\theta+i\gamma j}\ket{j}\, ,\quad S^+\ket{j}=-\frac{\sin(2\theta+\gamma j)}{\sin\gamma}\ket{j+1}\, ,\quad S^-\ket{j}=\frac{\sin(\gamma j)}{\sin\gamma}\ket{j-1}\, ,
\ee
where $j$ are integers. At roots of unity, the auxiliary space can be restricted to $j=0,\dots, m-1$, where $\gamma m\in \mathbb Z$, and, from the corresponding transfer matrices, it is possible to construct conservation laws which are quasi-local. 
We point out that the same procedure has been used to construct new quasi-local conservation laws also in other interacting models~\cite{qlcharges1}. 

In the noninteracting limit $\gamma=\frac{\pi}{2}$ (XX model), one obtains the entire set of conservation laws invariant under a one-site shift $\{I^{(n,+)},I^{(n,-)}\}$. 

What about the charges which are odd under a shift by one site?

\paragraph{Loop algebra.}%

At roots of unity $\gamma=\pi r$, the existence of non-commuting charges which are odd under a one-site shift is a consequence of an $sl_2$ loop algebra symmetry~\cite{DFM:2000}, like that underlying \eref{eq:loop}. However, so far, a non-abelian set of quasi-local charges has been constructed only for $\Delta=0$ \cite{F:super},
 and it is widely believed that for $\Delta\neq 0$ the extra charges are nonlocal. 

\paragraph{Additional quasi-local charges.} %

A feature which seems to be peculiar to interacting models is the presence of delocalized bound states. 
If interactions allow it, particles can bind together and form new species of excitations. This results in the appearance of additional families of conservation laws. In the XXZ model, these somehow correspond to (unitary) higher-spin representations of the quantum group algebra \eref{eq:qga}, namely
\begin{eqnarray}
\fl\qquad\qquad K\ket{j}=e^{i\gamma j}\ket{j}\qquad S^{\pm}\ket{j}=\frac{\sqrt{\sin(\gamma(S+1\pm j))\sin(\gamma(S\mp j))}}{\sin\gamma}\ket{j\pm 1}
\end{eqnarray}
where $S$ is half-integer or integer, and $j=-S,\dots,S$. 
By taking the derivatives of the logarithm of the transverse matrix at  the appropriate value of the spectral parameter, one obtains independent quasi-local conservation laws~\cite{newcharges, newGGE}. 

We refer the reader interested in the charges of the XXZ model to Review \cite{P:review} and references therein. 

\subsection{Currents} %

To the best of our knowledge, in the XXZ model the operator form of a generic current is  unknown. A brute-force calculation reveals that, in contrast to the XY model, the currents associated with reflection-symmetric charges are \emph{not} conserved. Apparently, there is only one exception, the energy current, which is equal to $Q_2$. 

\subsection{Expectation values in a macro-state}%

In the XXZ model, one can construct the excited states by acting with operators $B(\lambda_j)$ on a simple reference state, $\ket{\uparrow\cdots\uparrow}$.  Using the representation \eref{eq:tm}, these are given by 
\be
B(\lambda)=\mathrm{tr}_0\Bigl[\sigma_0^-\prod_{j=L}^1\mathcal L_j(\lambda)\Bigr]\, .
\ee
In order to be an excited state, the so-called rapidities $\lambda$ must satisfy the Bethe equations~\cite{Bethe}
\be
\Bigl[\frac{\sinh\left(\lambda_j+i\frac{\gamma}{2}\right)}{\sinh\left(\lambda_j+i\frac{\gamma}{2}\right)}\Bigr]^{L}
=\prod\nolimits_{\substack{l\neq j}}^N\Bigl[\frac{\sinh\left(\lambda_j-\lambda_l+i\gamma\right)}{\sinh\left(\lambda_j-\lambda_l-i\gamma\right)}\Bigr]\, .
\label{Eq:Bethe}
\ee
In the thermodynamic limit, the real parts of the rapidities become dense  and, exactly as discussed in \sref{ss:macro},  a thermodynamic state can be parametrized by a set of particle distributions $\{\rho_{j,{\rm p}}\}$, one for each string type. The excitations differ in the pattern of the solutions, which in the thermodynamic limit are organized in strings of rapidities with the same real part $\lambda^\alpha_j$ and equidistant imaginary parts~\cite{Takahashibook}. 
For example, for $\Delta=\frac{1}{2}$, \emph{i.e.} $\gamma=\frac{\pi}{3}$, there are three species of excitations:  the first corresponds to real solutions to the Bethe equations, the second  consists of  pairs of rapidities with imaginary part equal to $\pm \frac{\gamma}{2}$, and the third consists of one-strings with odd parity, \emph{i.e.}, with imaginary part equal to $\frac{\pi}{2}$. 

The expectation values of the charges have the form \eref{eq:root}, namely
\be\label{eq:QXXZ}
\braket{\rho|Q_\ell|\rho}=\sum_{n}\int{\rm d}\lambda\, q_n(\lambda) \rho_{n,{\rm p}}(\lambda)\, ,
\ee
for functions $q_n(\lambda) $ independent of the state. 
Remarkably, Refs \cite{CAD:hydro,BCNF} have recently found that the analogy to the noninteracting case extends to the currents, indeed one has  
\be\label{eq:JXXZ}
\braket{\rho|J_\ell[Q]|\rho}=\sum_{n}\int{\rm d}\lambda\, q_n(\lambda) v_n(\lambda)\rho_{n,{\rm p}}(\lambda)\, ,
\ee
where $v_n(\lambda)$ are the velocities of the excitations. Despite the formal equivalence between this expression and the noninteracting one \eref{eq:rhocurr}, there is a fundamental difference: in interacting models the velocities depend on the macro-state~\cite{bonnes14}!  

\section{Late-time dynamics after quantum quenches}\label{s:late}%

The late-time dynamics after quantum quenches $\rho_t=e^{-i H t}\rho_0 e^{i H t}$ is where all the elements introduced in the previous sections come to life. 

\subsection{Generalized Gibbs ensemble, representative state, and diagonal ensemble}\label{ss:GGE}%

We consider first the most common case where $\rho_0$ is a homogeneous state (like the ground state of some local Hamiltonian or a thermal state), and $H$ is translationally invariant. 
Contrary to what is expected of a quantum dynamics in a few particle system, in a  many-body quantum system like a spin chain, local observables can relax despite the system being closed.  Roughly speaking, this is because local observables have zero matrix elements  between macroscopically different states, and, in the thermodynamic limit, any information about the initial state which was not contained in the integrals of motion becomes eventually inaccessible. The state becomes locally equivalent to a macro-state. Describing the macro-state has been a central issue of the last decade~\cite{GGE0,CD:therm08,TFIC0,KSCCI,Pozsgay:13a,FE_13b,CE:reps,Collura_TG,nwbc-13,clusterGGE,Pozs_qB,FCEC_14,F:super,GGE_XXZ_Amst,XXZung,newGGE,Doyon,Eis_rev,quench_sinh,XXZGGEc, GGEbound}. 
We refer the reader interested in a comprehensive introduction to quench dynamics and relaxation in homogeneous isolated integrable quantum spin chains to Review~\cite{EF:review} and references therein. 
Here we just introduce the three most common representations of the late-time stationary state:
\begin{enumerate}
\item \emph{Generalized Gibbs ensemble}. This is based on the observation that the limit of infinite time corresponds to losing the ``maximal'' amount of information about the initial state. One can then represent the macro-state as the mixed state with maximal entropy, under the constraints of the relevant integrals of motion~\cite{EF:review,VR:review}:
\begin{eqnarray}
\rho^{\rm GGE}=\frac{1}{Z^{\rm GGE}}e^{-\sum_n\lambda_n Q^{(n)}}\nn
\{\lambda_j\}\ \mathrm{such\ that}\ \tr[\rho^{\rm GGE} Q_\ell^{(n)}]=\tr[\rho_0Q_\ell^{(n)}]\, .
\end{eqnarray}
Here $Q^{(n)}$ are the relevant charges, which generally are the local and quasi-local conservation laws.  
This can be seen as a generalization of the canonical description in statistical physics. 
\item \emph{Representative state}. This description relies on the fact that excited states with the same integrals of motion have the same local properties. Thus, one can pick one of them to represent the macro-state~\cite{CE:reps}. This is sometimes considered a generalization of the micro-canonical description in statistical physics.  
\item \emph{Diagonal ensemble}. This is based on the observation that the off-diagonal elements of the density matrix are characterized by  persistent oscillations in time, and hence must become irrelevant at late times, when the local observables become stationary.  The diagonal ensemble is then defined as follows:
\be\label{eq:DE}
\fl\qquad \rho^{\rm DE}=\lim_{L\rightarrow\infty}\lim_{t\rightarrow\infty}\frac{1}{t}\int_0^t\mathrm d\tau  e^{-iH \tau} \rho_0 e^{i H \tau}=\lim_{L\rightarrow\infty}\sum_{E_L}\braket{E_L|\rho_0|E_L}\ket{E_L}\bra{E_L}\, ,
\ee
where $\ket{E_L}$ is an orthonormal basis of steady states for the $L$-site chain.
\end{enumerate}
As long as the initial state is homogeneous and satisfies some basic properties, like cluster decomposition\footnote{A state $\rho$ has cluster decomposition properties if the correlation functions of generic localized observables $O_\ell$ and $O_\ell'$ satisfy
\be
\tr[\rho O_\ell O'_{\ell+r}]\overset{r\gg 1}{\longrightarrow}\tr[\rho O_\ell]\tr[ \rho O'_{\ell+r}]
\ee}, the three descriptions have always proved to be equivalent~\cite{genT,TFIC1,F:13,NLCE,GGE_XXZ_Amst}. 

\subsubsection{A controversial case.} %

A kind of exception to the aforementioned equivalence has been tacitly revealed in Refs \cite{Fdefect,CLopenXY}:
In the XY model in zero field, the set of the local conservation laws is non-abelian only if the number of sites is even; otherwise, the finite-chain counterparts of the additional charges $Y^{(n,s1,s2)}$ \eref{eq:QXY} do not exist (such charges are odd under a shift by one site),  and most of the exact degeneracy of the spectrum is removed. 
Strictly speaking, the thermodynamic limit in \eref{eq:DE} does not exist, and the equivalence with the other ensembles holds only taking sequences of chains with even size. 
This effect appears  if the initial state is not one-site shift invariant, for example, after quenches from states with antiferromagnetic order.
As a matter of fact, in such situations also the  initial state is sensitive to the parity of the chain's length, and  chains with an odd number of sites might be considered pathological.  
Nevertheless, such issues will return again, and get amplified, in the presence of a defect - \sref{s:loc} - where the situation will become practically unfixable. 

\subsection{Locally-quasi-stationary state}\label{ss:LQSS}%

In this section, we start relaxing the assumption of homogeneity. 
There is a vast literature also on this situation, and we refer the interested reader to the Reviews~\cite{BD:review, VM:review}, references therein, as well as to the references of \cite{BCNF, CAD:hydro}.

We start assuming that both the initial state and the (local) Hamiltonian are homogeneous in the region $\ell\geq \ell_{r}$, where $\ell_r=O(1)$.
Outside a light-cone propagating from $0$ at the Lieb-Robinson velocity $v^+_{LR}$, the form of the Hamiltonian in the region $\ell<\ell_r$ does not affect the late-time dynamics. Indeed, using the Lieb-Robinson bound~\cite{LR72} and the results of \cite{bravyi06}, one can easily show~\cite{BF:16}
\be
\tr\bigl[\rho_0 e^{i H t} O_\ell  e^{-i H t}\bigr]\overset{\ell\gtrsim v_{LR}^+ t\atop t\rightarrow\infty}\longrightarrow  \tr\bigl[\rho_0 e^{i  H_+ t} O_\ell  e^{-i  H_+ t}\bigr]\, ,
\ee
where $H_+$ is a translationally invariant effective Hamiltonian obtained by extending the known part of the Hamiltonian over the entire chain.

Analogously, any information about the state in the region $\ell<\ell_r$ turns out to be irrelevant. Indeed, the observable $e^{i  H_+ t} O_\ell  e^{-i  H_+ t}$ is exponentially well approximated (in the time) by an operator with support in $\ell\geq \ell_r$ for any fixed $\frac{\ell}{t}>v_{LR}^+$ \cite{bravyi06}. As a result, we can write
 \be\label{eq:step2}
\tr\bigl[\rho_0 e^{i H_+ t} O_\ell  e^{-i H_+ t}\bigr]\overset{\ell\gtrsim v^+_{LR} t\atop t\rightarrow\infty}\longrightarrow  \tr\bigl[\rho_{0}^+ e^{i  H_+ t} O_\ell  e^{-i  H_+ t}\bigr]\, ,
\ee
where $\rho_{0}^+$ is an effective initial state obtained by assuming homogeneity and extending the known part of the state ($\ell\geq \ell_r$) over the entire chain. 
For the right hand side of \eref{eq:step2}, we can use the results of the previous subsection; we finally obtain
\be\label{eq:BCR}
\tr\bigl[\rho_0 e^{i H t} O_\ell  e^{-i H t}\bigr]\overset{\ell\gtrsim v_{LR}^+ t\atop t\rightarrow\infty}\longrightarrow \tr\bigl[\rho^{{\rm GGE}_+} O_\ell \bigr]\, ,
\ee
where $\rho^{{\rm GGE}_+}$ represents the macro-state associated with the  effective homogeneous quench $\rho_{t}^+=e^{-i  H_+ t}\rho_{0}^+ e^{i  H_+ t}$. 

Almost irrespective of the specific properties of the state and of the Hamiltonian in the region $\ell<\ell_r$, the dephasing mechanisms underlying relaxation are active also inside the light-cone $0\lesssim \frac{\ell}{t}\lesssim v_{LR}^+ $; one expects the expectation values of local observables to be described by a position-dependent macro-state.
This is determined by the expectation values of the \mbox{(quasi-)local} conservation laws, so the continuity equation \eref{eq:continuity} for the charges is the main tool to connect different macro-states inside the light-cone. 

As long as $\ell\gg \ell_r$, the continuity equations are written in terms of  charges and currents of the effective Hamiltonian $H_+$. In particular, if $H_+$ describes an integrable model, the quasi-particle excitations move ballistically and, at late times, charges and currents are expected to become functions of the ray $\zeta=\frac{\ell}{t}$. 
We will refer to the resulting ray-dependent macro-state as LQSS (locally-quasi-stationary state)~\cite{BF:16}.

If the expectation values of charges and currents are sufficiently smooth functions of $\zeta$ in some interval  $(\zeta_l,\zeta_r)$, the LQSS in $(\zeta_l,\zeta_r)$ is completely characterized by 
\begin{itemize}
\item[-] the continuity equations \eref{eq:continuity}, which, in the limit considered, can be rewritten as
\be
\partial_t \tr\bigl[\rho^{\rm LQSS}_{\ell/t} Q_\ell^{+(n)}\bigr]+\partial_x\tr\bigl[\rho^{\rm LQSS}_{\ell/t}J_\ell[Q^{+(n)]}\bigr]=0\, ;
\ee
\item[-] the states at the boundaries, \emph{i.e.} $\rho^{\rm LQSS}_{\zeta_l}$ and $\rho^{\rm LQSS}_{\zeta_r}$.
\end{itemize}
This claim is based on the representation of the expectation values through the densities of excitations, namely \eref{eq:rhocurr} for noninteracting models and the analogous ones, \eref{eq:QXXZ} and \eref{eq:JXXZ}, for interacting models like the XXZ spin-$\frac{1}{2}$ chain. Indeed, we have
\be
\sum_n\int\mathrm d \lambda q^+_n(\lambda) [\partial_t \rho^+_{n,{\rm p}}(\lambda,x,t)+\partial_x (v_n^+(\lambda,x,t) \rho^+_{n,{\rm p}}(\lambda,x,t))]=0\, ,
\ee
where $\rho^+_{n,{\rm p}}(\lambda,x,t)$ are the densities of the particle excitations of $H^+$, $v_n^+(\lambda,x,t)$ are the corresponding velocities, and $q^+_n(\lambda)$ are the single-particle eigenvalues, which are functions, independent of the state, characterizing the (quasi-)local conservation laws. 
If $q_n(\lambda)$ form a complete set of functions this implies
\be\label{eq:contzeta}
\zeta\partial_\zeta \rho_{n,{\rm p}}^+(\lambda,\zeta)-\partial_\zeta(v_n^+(\lambda,\zeta) \rho_{n,{\rm p}}^+(\lambda,\zeta))=0\, ,
\ee
where we assumed that the densities depend on position and time only through the ratio $\zeta=x/t$.

In noninteracting models, the velocities $v_n^+(\lambda,\zeta)$ are independent of $\zeta$, and the solution $\rho_{n,{\rm p}}^+(\lambda,\zeta)$ to the continuity equation is a piecewise constant function of $\zeta$, which can be discontinuous at $\zeta=v_n^+(\lambda)$.  For given $\lambda$, in $(\zeta_l,\zeta_r)$ this equation can have one solution at the most, so the solution to \eref{eq:contzeta} depends only on the values of the root densities at the boundaries. 

In interacting integrable models the velocity depends on $\zeta$, nevertheless, in the XXZ spin-$\frac{1}{2}$ chain, Ref.~\cite{BCNF} has shown that  the solution can be recast in the same form as in the noninteracting case, provided to replace the root densities by the so-called filling functions $\vartheta_n(\lambda,\zeta)$, which are the particle densities for given momentum of the excitation (they are the analogues of the occupation numbers in interacting models)\footnote{The filling functions differ from $\rho_{n,{\rm p}}(\lambda,\zeta)$ only in the Jacobian of the transformation between rapidities $\lambda$ and momenta $p_n(\lambda)$: $\vartheta_n(\lambda,\zeta)=\frac{\rho_{n,{\rm p}}(\lambda,\zeta)}{\rho_{n,{\rm tot}}(\lambda,\zeta)}$, where  $\rho_{n,{\rm tot}}(\lambda,\zeta) \mathrm d\lambda=\mathrm d p_n(\lambda)$~\cite{Korepinbook}.
}. 
Provided that, for given $\lambda$, the equation $\zeta=v_n^+(\lambda,\zeta) $ does not have more than one solution in $(\zeta_l,\zeta_r)$, the solution to \eref{eq:contzeta} depends only on the values of the root densities at the boundaries.  

Going back to our problem, we have already identified one boundary condition, namely \eref{eq:BCR}, which corresponds to $\zeta_r>v_{LR}^+$. In order to find the LQSS, we need to determine the other boundary condition(s). 

In the simple case $H=H_+$, if the initial state is homogeneous also in the region $\ell<\ell_l$ (for a given $\ell_l<\ell_r$) we can apply the previous procedure in reverse, which gives 
\be\label{eq:BCL}
\tr\bigl[\rho_0 e^{i H t} O_\ell  e^{-i H t}\bigr]\overset{\ell\lesssim -v^-_{LR} t\atop t\rightarrow\infty}\longrightarrow \tr\bigl[\rho^{{\rm GGE}_-} O_\ell \bigr]\, .
\ee
Here $\rho^{{\rm GGE}_-}$ represents the macro-state associated with the  effective homogeneous quench $\rho_{t}^-=e^{-i  H_+ t}\rho_{0}^- e^{i  H_+ t}$, and $\rho_{0}^-$ is an effective initial state obtained by assuming homogeneity and extending the left part of the state ($\ell\leq \ell_l$) over the entire chain.
This situation has been studied in Ref.~\cite{BCNF} for quenches in the XXZ model starting from states consisting of two parts joined together. In all the cases investigated (two chains prepared at different temperatures, in the ground states of different Hamiltonians, in the ``domain-wall state''), numerical simulations have confirmed the validity of the procedure. 

\begin{figure}
\begin{center}

\begin{tikzpicture}[xscale=2.4, yscale=1.45]

\fill[black!0] (0,0) -- (2.3,0) --(2.3, 2.4) -- (0,2.4);
\fill[black!10] (0,0) -- (-1.8,0) --(-1.8, 2.4) -- (0,2.4);
\fill[green!60!blue!70] (0,0) -- (0.6,2.4) -- (0, 2.4) -- (0,0);
\fill[orange!60!blue!70] (0,0) -- (-0.8,2.4) -- (0, 2.4) -- (0,0);
\draw[-, line width=0.8mm,white] (0,0) -- (0,2.4);
\draw[yellow,thick] (0,0) -- (0.6*1.05,2.4*1.05);
\draw[dashed] (0,0) -- (0.6*1.05,2.4*1.05);
\draw node [above] at (0.6*1.05,2.5) {$\zeta_\lambda^+=v_n^+(\lambda,\zeta_\lambda^+)$};

\draw[yellow,thick] (0,0) -- (-0.8*1.05,2.4*1.05);
\draw[dashed] (0,0) -- (-0.8*1.05,2.4*1.05);
\draw node [above] at (-0.8*1.05,2.5) {$\zeta_\lambda^-=v_n^-(\lambda,\zeta_\lambda^-)$};

\fill[white,path fading=fade right] (0,0) -- (0.1,0) --(0.1, 2.4) -- (0,2.4);
\fill[white,path fading=fade left] (-0.13,0) -- (0,0) --(0, 2.4) -- (-0.13,2.4);

\draw[-] (0,0) -- (1*1.25,2.5);
\draw[-] (0,0) -- (-1.2*1.25,2.5);

\draw node[below] at (1.6,0) {$\ell$};
\draw node[below] at (0,0) {$0$};
\draw node[below] at (0.13,0) {$\ell_r$};
\draw[-, line width=0.2mm] (0.1,-0.05) -- (0.1,0) ;
\draw node[below] at (-0.13,0) {$\ell_l$};
\draw[-, line width=0.2mm] (-0.135,-0.05) -- (-0.135,0) ;
\draw node[left] at (0, 2.55) {$t$};

\draw[->, line width=0.4mm] (-1.8,0) -- (1.8,0) ;
\draw[-, line width=0.4mm,brown!60] (-1.8,0) -- (0,0) ;
\draw[->, line width=0.4mm] (0,-0.05) -- (0,2.7);
\draw node[left] at (1.6,0.8) {$\vartheta_{n}^+(\lambda,v^+_{LR})$};
\draw node[right] at (-1.6,0.8) {$\vartheta_{n}^-(\lambda,-v^-_{LR})$};
\draw[yellow,font=\relsize{2}] node[below] at (0.2,2) {?};
\draw[yellow,font=\relsize{2}] node[below] at (-0.3,2) {?};
\draw[font=\relsize{2}] node[left] at (1.8,1.6) {$\mathrm{GGE}_+$};
\draw[font=\relsize{2}] node[right] at (-1.8,1.6) {$\mathrm{GGE}_-$};
\draw[dotted, line width=0.5mm,white] (-0.13,0) -- (0.1,0) ;
\end{tikzpicture}

\caption{Cartoon of the light-cone corresponding to the most general case considered in \sref{ss:LQSS}: the initial state and the Hamiltonian are homogeneous in the bulk on the right ($\ell\gg\ell_r$) and on the left ($\ell\ll\ell_l$). 
At fixed rapidity $\lambda$, the filling functions $\vartheta_n(\lambda,\ell/t)$ are piecewise functions of $\ell/t$ with three discontinuities at the most: one at $\ell/t\rightarrow 0$, corresponding to the region where the form of the Hamiltonian is not specified, and two at the solutions $\ell/t=\zeta^\pm_\lambda$ to the equations $\pm \zeta^\pm_\lambda=v_n^\pm (\lambda,\pm \zeta^\pm_\lambda)\in [0,v^\pm_{LR}]$. The solid lines represent the light-cone. 
The regions marked with a question mark depend on the form of the Hamiltonian close to site $0$. 
If the Hamiltonian is homogeneous, only one discontinuity is present (three consecutive colors must be identified). }\label{fig:0}
\end{center}
\end{figure}
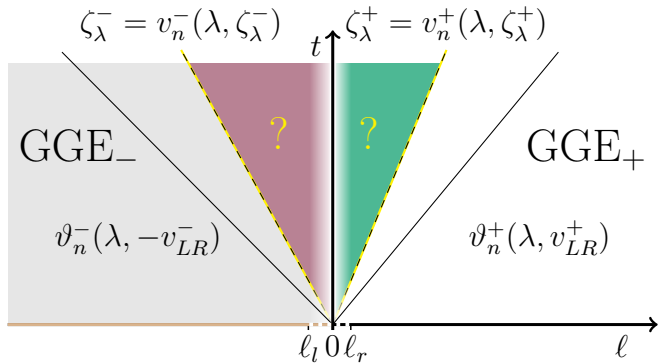

By relaxing the assumption of homogeneity of the Hamiltonian, the continuity equation for $H_+$ is not sufficient anymore to connect the LQSS with $\ell>\ell_r$ to the LQSS  with $\ell<\ell_r$ - \fref{fig:0}. 
Thus, it becomes necessary to find a complementary procedure to 
determine the late-time dynamics ``close'' to the defect.  

We will restrict to the simplest type of Hamiltonian inhomogeneity, where $H_+$ differs from $H$ only in a localized defect with support in $[\ell_l,\ell_r]$.  We mention that a case where the Hamiltonian to the left is different from the Hamiltonian to the right was considered in Ref.~\cite{2Delta}. 

\section{Localized defects}\label{s:loc}%

In this section, we investigate the effects of a localized Hamiltonian defect, first, on the conservation laws, and, second, on the late-time dynamics. 
We will keep the discussion very general, and use the models introduced in the first part of the paper to construct explicit examples and counterexamples.

\subsection{On the conservation laws}%

We write the Hamiltonian as follows
\be
H^{[{\mathtt s}]}=H^{[0]}+\mathtt s d\, ,
\ee
where $H^{[0]}$ is translationally invariant, $d$ is a (quasi-)localized defect, and $\mathtt s=0,1$. We indicate a generic (quasi-)local charge of $H^{[0]}$ by $Q^{[0]}$.
We can distinguish two cases~\cite{Fdefect}:
\begin{itemize}
\item The charge $Q^{[0]}$ is \emph{deformed} by the defect, that is to say, there is a (quasi-)localized operator $\delta Q$ (its operator norm is finite $\parallel \delta Q\parallel<\infty$) such that $[H^{[1]},\delta Q]=[Q^{[0]},d]$. Thus, $Q^{[0]}+\delta Q$ is (quasi-)local and commutes with $H^{[1]}$.
\item The charge  $Q^{[0]}$ becomes \emph{extinct}, that is to say, no (quasi-)localized deformation is sufficient to zero the commutator with $H^{[1]}$. 
\end{itemize}
As a matter of fact, there is a third option, which is a special subclass of extinct charges associated with conserved operators in the presence of the defect:
\begin{itemize}
\item \emph{Crossing-over}. Two charges $Q^{[0](1)}$ and $Q^{[0](2)}$ of $H^{[0]}$ \emph{hybridize} to form a \mbox{(quasi-)local} charge of $H^{[1]}$. That is to say, there is a (quasi-)localized operator $\delta Q$ such that $[H^{[1]},\delta Q]=i (J^{[0](2)}_{R}-J^{[0](1)}_{L})$, where $J^{[0](j)}_{L/R}$ is the density current of the charge $Q^{[0](j)}$ at the left and at the right hand side of the defect, respectively. In other words, there is a charge of  $H^{[1]}$ which is equivalent to $Q^{[0](1)}$ ($Q^{[0](2)}$) in the bulk on the left (right) hand side of the defect. 
\end{itemize}

We point out that, in principle, there could be also conservation laws of $H^{[1]}$ associated with the trivial charge of $H^{[0]}$, \emph{i.e.}, the identity. These are quasi-localized around the defect and are usually identified as ``zero modes''. It is very common to find such charges in noninteracting spin chains~\cite{Kitaev,CLopenXY}, and they have been recently found also in interacting models when the defect switches off the interaction between two sites~\cite{Fend}. We will assume that zero modes are not produced by the defect, which is generally expected if the ground state of $H^{[1]}$ is \emph{not} in an ordered phase.

\subsubsection{Defect current.}\label{ss:current}   %
We define the ``\emph{anomalous defect current}'' as follows
\be
\mathcal A_d[{Q^{[0]}},\rho_0]=\lim_{t\rightarrow\infty}\tr\bigl(e^{i H^{[1]} t} \rho_0 e^{-i H^{[1]} t}i [Q^{[0]},d]\bigr)\, ,
\ee
where we assumed that the limit of infinite time exists. 
If $Q^{[0]}$ is (quasi-)local, this is the late-time expectation value of a (quasi-)local operator after a quench from $\rho_0$. The reason why we qualify it as ``anomalous'' will be clear before later. 
 
\subsubsection{Deformed charges.}\label{ss:deformed}  %

To give an example of deformed charges, we can consider a defect of the form
\be
d=e^{iV} H^{[0]} e^{-i V}-H^{[0]}\, ,
\ee
where $V$ is (quasi-)localized around a given position. Since this corresponds to the unitary transformation $H^{[1]}=e^{iV} H^{[0]} e^{-i V}$, the charges are mapped into $Q^{[1]}=e^{iV} Q^{[0]} e^{-i V}$ and are different from the charges of the unperturbed Hamiltonian only for terms (quasi-)localized around the defect. 

Another interesting class of defects  is the following
\be\label{eq:defc}
\fl\qquad d=e^{i\theta\sum_{\ell>0}Q^{[0]}_\ell}H^{[0]}e^{-i\theta\sum_{\ell>0}Q^{[0]}_\ell}-H^{[0]}=\int_0^\theta\mathrm d s e^{i s\sum_{\ell>0}Q^{[0]}_\ell}J^{[0]}_0[Q^{[0]}]e^{-i s\sum_{\ell>0}Q^{[0]}_\ell}\, ,
\ee
where $Q^{[0]}_\ell$ is the density of a local charge 
\be
Q^{[0]}=\sum_\ell Q^{[0]}_\ell\, ,
\ee
and $J^{[0]}_0[Q^{[0]}]$ is the associated current density without the defect
\be
J^{[0]}_0[Q^{[0]}]=i [\sum_{\ell>0}Q^{[0]}_\ell,H^{[0]}]\, .
\ee
Also in this case, any charge commuting with $Q^{[0]}$ is deformed by the defect. More generally, there can be deformed charges even if the Hamiltonian with the defect is not unitarily equivalent to $H^{[0]}$. A simple example is given by the transverse field-Ising chain ($\gamma=1$ in \eref{eq:XY}), with a defect that cuts the interaction between two sites: the reflection symmetric charges turn out to be deformed~\cite{CLopenXY}.

As shown in Ref.~\cite{Fdefect}, deformed charges do not have an anomalous defect current, \emph{i.e.}, $\mathcal A_{d}[Q^{[0]},\rho_0]=0$.  
Indeed, the existence of a quasi-local deformation $\delta Q$ implies
\be
\fl\qquad \bigl|\tr\bigl[\rho_0(Q^{[0]}-e^{i H^{[1]}t} Q^{[0]}e^{-i H^{[1]}t})\bigr]\bigr|= \bigl|\tr\bigl[\rho_0(e^{i H^{[1]}t}\delta Qe^{-i H^{[1]}t}-\delta Q)\bigr]\bigr|\leq 2\parallel\delta Q\parallel\, ;
\ee
on the other hand, the left hand side reads as
\be
\fl\frac{1}{t}\tr\bigl[\rho_0(Q^{[0]}-e^{i H^{[1]}t} Q^{[0]}e^{-i H^{[1]}t})\bigr]=\int_0^t\frac{\mathrm d \tau}{t} \tr\bigl[\rho_0e^{i H^{[1]}\tau}i[Q^{[0]},d]e^{-i H^{[1]}\tau}\bigr]\overset{t\rightarrow\infty}{\longrightarrow}\mathcal A_{d}[Q^{[0]},\rho_0].
\ee
Since $\parallel\delta Q\parallel$ is finite, $\mathcal A_{d}[Q^{[0]},\rho_0]$ must be zero. 

\subsubsection{Extinct charges.}\label{ss:extinct}  %

Like for the deformed charges, also the appearance of extinct charges can be understood by switching on a defect that results in a unitary transformation. 

In the XY model in zero field, we can exploit the nontrivial algebra of the local conservation laws to design a defect, of the form \eref{eq:defc}, which, however, results in the extinction (hybridization) of infinitely many charges. 
For example, let us consider a defect proportional to the density current $J_0[Y^{(0,+-)}]$, where $Y^{(0,+-)}$ is one of the local charges that are odd under a shift by one site (\emph{cf}. \eref{eq:QXY}).
In the limit of small coupling constant, this defect is equivalent to the unitary transformation \eref{eq:defc} with $Q^{[0]}=Y^{(0,+-)}$. Since $Y^{(0,+-)}$ does not commute with $I^{(2n,-)}$ and $I^{(2n+1,+)}$ (the corresponding symbols, reported in \ref{a:nonint}, do not commute), such charges will  hybridize with charges that are not one-site shift invariant. 

For a more standard example, let us consider the XXZ model. The Hamiltonian is invariant under spin-flip $\Pi^x$, but, as discussed in \sref{ss:chargeXXZ}, especially for $\Delta=\cos(\pi r)$ (with $r$ a rational number), there are charges which are odd under that transformation. Let us then act with spin-flip only on half chain
\be\label{eq:def}
d=\Bigl(\prod_{\ell>0}\sigma_\ell^x\Bigr) H^{XXZ}_\Delta \Bigl(\prod_{\ell>0}\sigma_\ell^x\Bigr)-H_\Delta^{\rm XXZ}=-\frac{{\mathrm J}}{2}(\sigma_0^y\sigma_1^y+\Delta \sigma_0^z\sigma_1^z)\, .
\ee
This preserves integrability, and also the spectrum of the Hamiltonian. On the other hand, the charges transform as
\be
Q\rightarrow \Bigl(\prod_{\ell>0}\sigma_\ell^x\Bigr)  Q \Bigl(\prod_{\ell>0}\sigma_\ell^x\Bigr) \, .
\ee
Consequently,  the conservation laws of the XXZ model which are invariant under spin flip are deformed by the defect, whereas the odd conservation laws (\emph{e.g.} $S^z$) are mapped to hybrid charges whose densities have different signs on the left and on the right hand side of the defect.
For the sake of clarity, we stress that extinct charges do not appear only when $H^{[1]}$ and $H^{[0]}$ are unitarily equivalent. For example, in the transverse field-Ising chain, all the antireflection symmetric charges are destroyed by a defect that cuts the interaction between two sites~\cite{CLopenXY}.

In contrast to the deformed charges, the anomalous defect current of an extinct charge can be different from zero. Let us consider again the previous example.  As extinct charge, we choose the magnetization $S^z$, which is odd under spin flip. We have
\be
i[S^z,d]=-\frac{{\mathrm J}}{2}(\sigma_0^y\sigma_1^x+\sigma_0^x\sigma_1^y)=2\sigma_1^xJ^{[0]}_0[S^z]\sigma_1^x\, ,
\ee
where $J^{[0]}_0[S^z]$ is the unperturbed spin density current at position $0$. 
We follow the time evolution of $i[S^z,d]$ starting from the state $\ket{\Uparrow}$ with all spins up
\be\label{eq:DW}
\fl \qquad \braket{\Uparrow|e^{i (H^{\rm XXZ}_\Delta+d) t}i [S^z,d]e^{-i (H^{\rm XXZ}_\Delta+d) t}|\Uparrow}=2\braket{\Uparrow\Downarrow | e^{i H^{\rm XXZ}_\Delta t}  
 J_0[S^z]
e^{-i H^{\rm XXZ}_\Delta t}|\Uparrow\Downarrow}\, .
\ee
Here we used  the explicit form of the defect \eref{eq:def}, and $\ket{\Uparrow\Downarrow}$ is the domain-wall state with all spins up until site $0$ and all spins down from site $1$. 
The right-hand side is well-known to remain  nonzero even at infinite times for $\Delta=\cos(\pi r)$. For example, for $\Delta=\frac{1}{2}$,  it approaches $\mathcal A_{S^z}[\ket{\Uparrow}\bra{\Uparrow}]\approx 0.477{\mathrm J}$. 
This number has been computed in Ref.~\cite{BCNF} by solving the continuity equation \eref{eq:contzeta} with the boundary conditions $\vartheta_{j}(\lambda,-{\mathrm J})=0$ (corresponding to the excited state $\ket{\Uparrow}$) and $\vartheta_{j}(\lambda,{\mathrm J})=1-\delta_{j 1}$ (corresponding to the excited state $\ket{\Downarrow}$), where $j=1,2,3$ labels the three species of excitations of the model. 
In conclusion, \emph{there is an anomalous defect current associated with the extinct (hybridized) charge $S^z$}.
We now show that this has striking consequences. 

\subsection{Late-time dynamics: failure of the diagonal ensemble.}%

In the limit of infinite time, one generally expects that the stationary behavior of observables can be described by a stationary state. A naive explanation is that a density matrix describing the stationary expectation values of the observables  should commute with the Hamiltonian, or some time dependence would remain. 

For the sake of concreteness, we focus again on the example of the previous subsection, but any conclusion will hold true in any other situation with a nonzero anomalous defect current. 
The commutator $i [S^z,d]$ is localized; one is tempted to  replace  the time evolving state  by an ensemble commuting with the Hamiltonian:
\begin{eqnarray}
\mathcal A_{S^z}[\ket{\Uparrow}\bra{\Uparrow}]=\lim_{t\rightarrow\infty}\braket{\Uparrow|e^{i H^{[1]} t}i [S^z,d]e^{-i (H^{\rm XXZ}_\Delta+d) t}|\Uparrow}\overset{?}{=} \mathrm{tr}[\bar \rho\,  i[S^z,d]] \\
{}[\bar \rho,H^{\rm XXZ}_\Delta+d]=0\, .
\end{eqnarray}
On the other hand, we have
\be\label{eq:paradox}
\mathrm{tr}[\bar \rho\,  i[S^z,d]]=\mathrm{tr}[\bar \rho\,  i[S^z,H^{\rm XXZ}_\Delta +d]]=\mathrm{tr}[i [H^{\rm XXZ}_\Delta +d,\bar \rho]S^z]=0\, ,
\ee
that is to say, \emph{the anomalous defect current is zero}!? 
This is clearly in contradiction with the spin current being nonzero after quenching from the domain-wall state \eref{eq:DW}. 
The apparent inconsistency is a manifestation of the presence of almost but not exactly degenerate excited states that are connected by local observables. 
In the thermodynamic limit, such states become degenerate, and the paradox is resolved. However, 
there is no possibility to circumvent the absurd working in finite chains, and \emph{any} attempt to describe the expectation value by means of a stationary state is doomed to fail. This is independent of whether the stationary state is chosen to be a Gibbs, a generalized Gibbs ensemble, or a representative state. The expectation value of $i[S^z,d]$ in any excited state of $H^{\rm XXZ}_\Delta +d$ is exactly zero. 

In all the descriptions mentioned so far (GE, GGE, representative state),  \eref{eq:paradox} is not a real issue because the thermodynamic limit should be taken before the infinite time limit. On the other hand, the paradox undermines the description in terms of  the diagonal ensemble~\eref{eq:DE},  which is instead based on the infinite time average in finite chains.  
Because of \eref{eq:paradox}, $\rho^{\rm DE}$ can not describe a nonzero anomalous defect current.  

In conclusion, we have shown that a \emph{localized defect is sufficient to lead the diagonal ensemble to failure}. 

It is worth pointing out that, in simple cases, the paradox can be resolved by adding a defect at infinity. For example, a finite-chain analogue of the case studied before with the defect \eref{eq:def} consists of \emph{two} defects, $d_1$ and $d_{L/2}$ (the string of $\sigma^x$ extends over half chain), one of which is sent to infinity in the thermodynamic limit. 
Then, \eref{eq:paradox} only tells us that the anomalous defect current associated with one defect is neutralized by the   anomalous defect current of the other defect. 
While the details of the finite-volume analogue of the model play a key role in the construction of the diagonal ensemble, in the thermodynamic limit (but also in the finite chain if the time is sufficiently smaller than the time needed to traverse the chain), time evolution is not affected by the details of the Hamiltonian at infinity.  

\subsection{Generic localized defects}\label{ss:genloc}%

So far, we have shown the paradox underlying \eref{eq:paradox} only in integrable models perturbed with defects that preserve integrability. A crucial question is whether this kind of behavior emerges also for generic defects.
It is widely believed that a generic localized defect is responsible for the breaking of integrability. 
Let us then assume that the only quasi-local charge commuting with $H$, up to boundary terms (which are sent to infinity in the thermodynamic limit), is the Hamiltonian itself. 

 Does it mean that the stationary values of the observables close to the defect can be described by a thermal ensemble? 
 
 First, we remind the reader that, in one dimensional systems at temperature different from zero, long-range order is generally forbidden~\cite{Mermin,Hohenberg}; the details of the Hamiltonian far away from the support of an observable are not expected to modify its thermal expectation value (for sufficiently high temperature, this statement has been proven  in any dimension in Ref.~\cite{localT}). 
As a result, it is reasonable to expect that the sequence of the expectation values of a given observable $O$ in thermal states of chains with increasing lengths approaches the thermal expectation value in the thermodynamic limit
\be\label{eq:localT}
\lim_{L\rightarrow\infty}\frac{\tr[e^{-\beta H_L} O]}{\tr[e^{-\beta H_L}]}=\frac{\tr[e^{-\beta H_\infty} O]}{\tr[e^{-\beta H_\infty}]}\, .
\ee
Here $H_L$ is the Hamiltonian in a periodic chain with $L$ sites, and $H_\infty$ is the Hamiltonian in the thermodynamic limit.

Not having solid reasons to question \eref{eq:localT}, we shall assume its validity. 

By combining \eref{eq:paradox} with \eref{eq:localT}, we deduce that, in thermal states, the anomalous defect current of any \mbox{(quasi-)local} $Q^{[0]}$ commuting with the unperturbed  Hamiltonian is zero, \emph{i.e.}, \emph{a nonzero anomalous defect current is incompatible with thermalization}.  

Let us then inspect the anomalous defect current in the presence of generic defects acting on one or two sites. We shall restrict ourselves to defects of the form
\be\label{eq:defect}
d=\frac{1}{2}\vec n_0\cdot\vec\sigma_0+\frac{1}{2}\vec n_1\cdot\vec\sigma_1\, ,
\ee
for arbitrary values of $\vec n_0$ and $\vec n_1$, and study the time evolution of the state $\ket{\Uparrow}$ with all spins up under a perturbed XXZ Hamiltonian \eref{eq:HXXZ} with $\Delta=\frac{1}{2}$. The initial state is an excited state of the XXZ model, so the evolution is nontrivial only inside the light-cone described in \sref{ss:LQSS}. This choice is numerically convenient, allowing to simulate rather large times, even using exact diagonalization techniques. 

\begin{figure}
\begin{center}\includegraphics[width=0.48\textwidth]{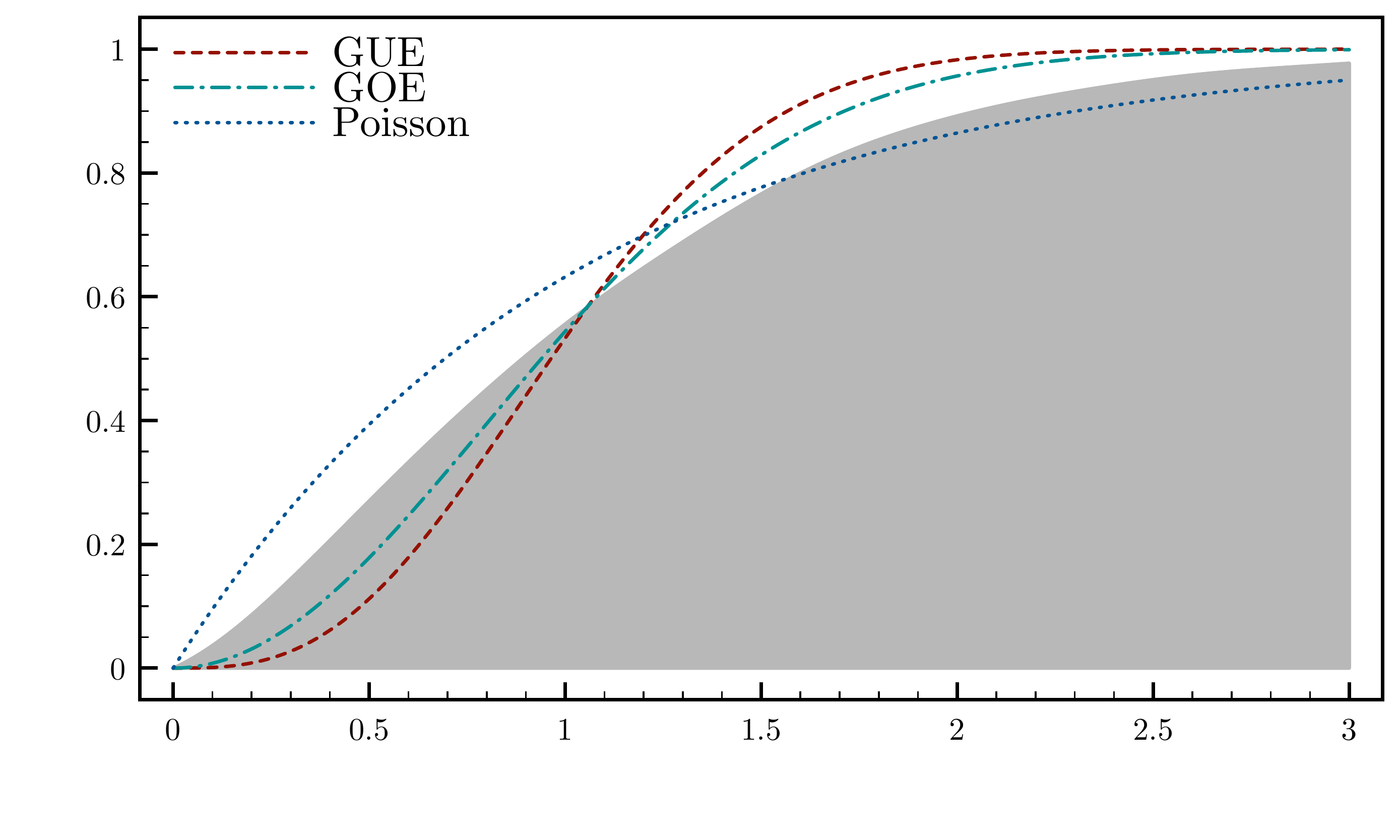}\hspace{0.5cm}
\includegraphics[width=0.48\textwidth]{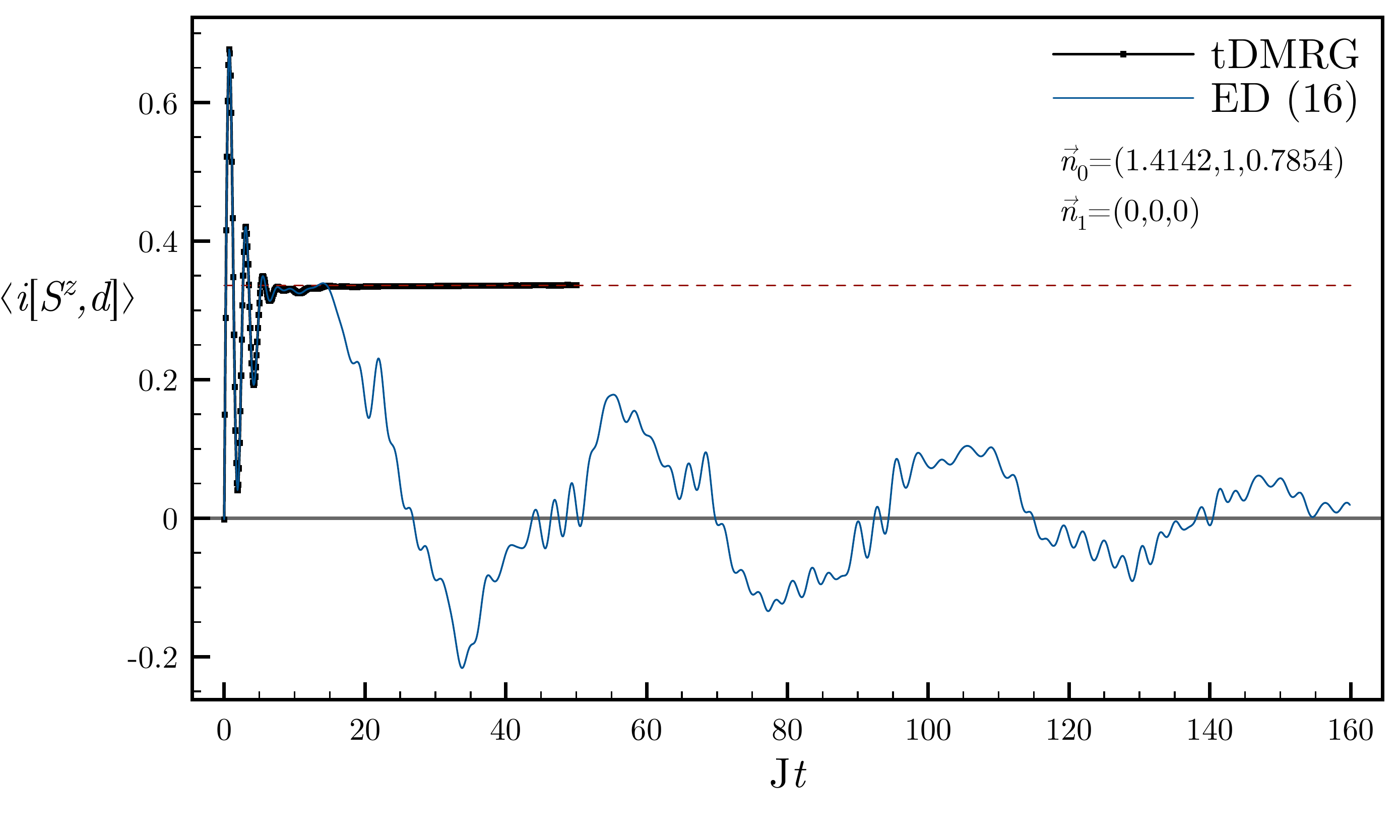}
\caption{Left: The filled region represents the cumulative energy level spacing distribution for the 1024 excited states with the energy closest to the energy of the state with all spins up $\ket{\Uparrow}$ in a chain with $14$ spins. GUE and GOE stand for Gaussian unitary ensemble and Gaussian orthogonal ensemble, respectively.  Right: Time evolution of $i[S^z,d]$ after a quench from $\ket{\Uparrow}$ under the XXZ Hamiltonian with $\Delta=\frac{1}{2}$ and the defect \eref{en:def1}. The  thick black curve with $\tiny\fullsquare$ is the result of a tDMRG simulation~\cite{Mario} in a chain with $100$ sites; the estimated errors are smaller than the symbol's size.
The blue curve is the time evolution in a chain with $16$ spins.  
The dashed red horizontal line is a guide to the eye. }\label{fig:1}
\end{center}
\end{figure}

\begin{figure}
\begin{center}
\includegraphics[width=0.48\textwidth]{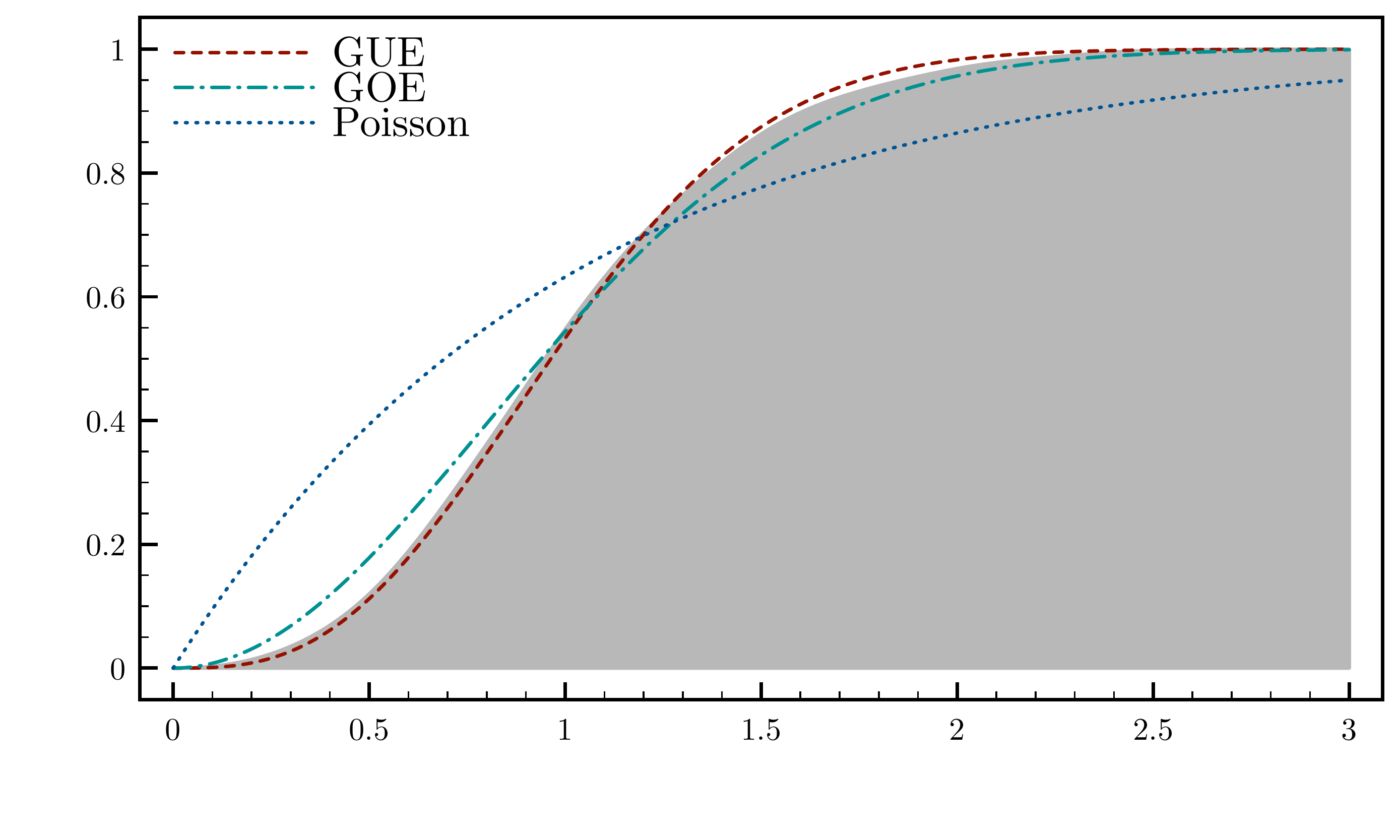}\hspace{0.5cm}
\includegraphics[width=0.48\textwidth]{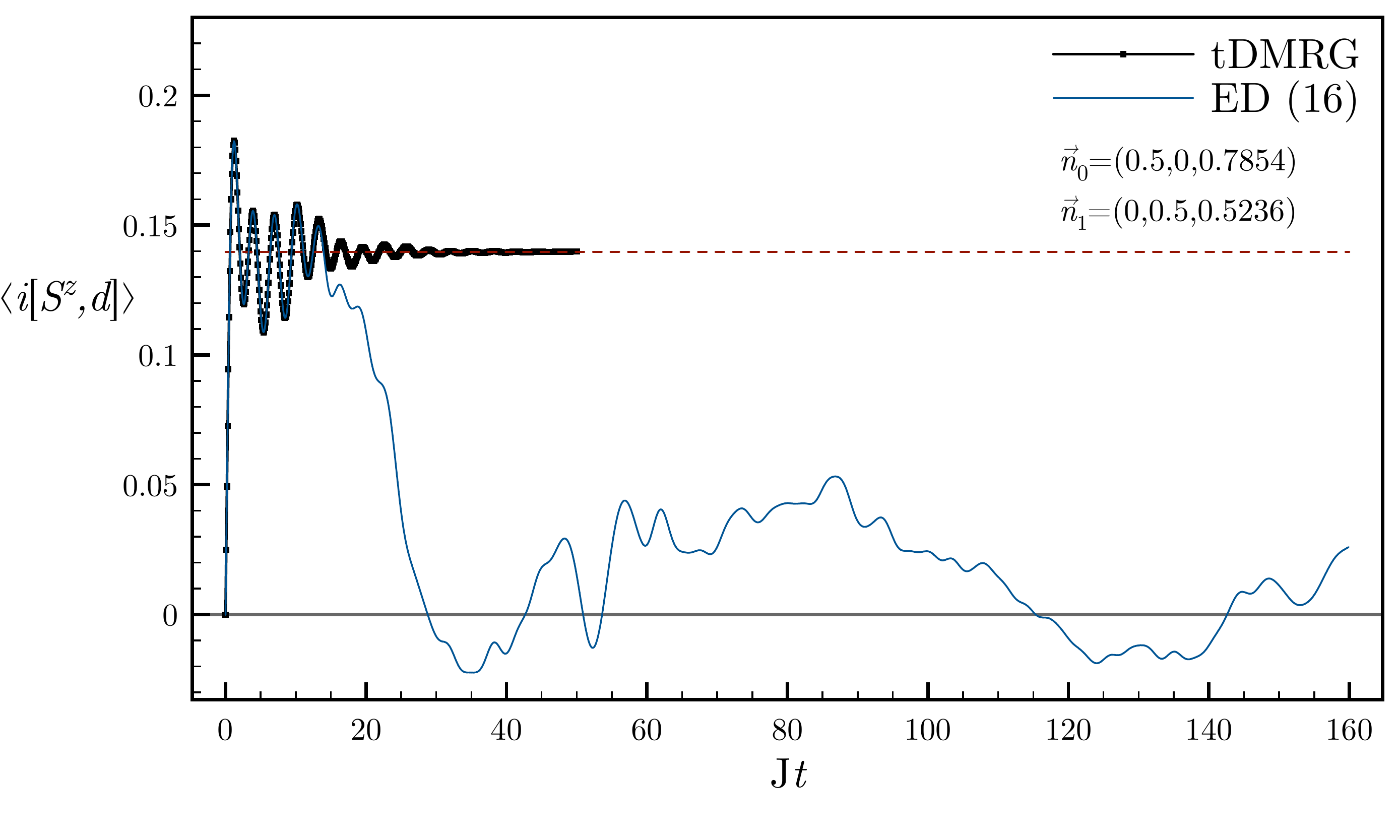}
\caption{The same as in \fref{fig:1} for the defect \eref{en:def2}. }\label{fig:2}
\end{center}
\end{figure}

\begin{figure}
\begin{center}
\includegraphics[width=0.48\textwidth]{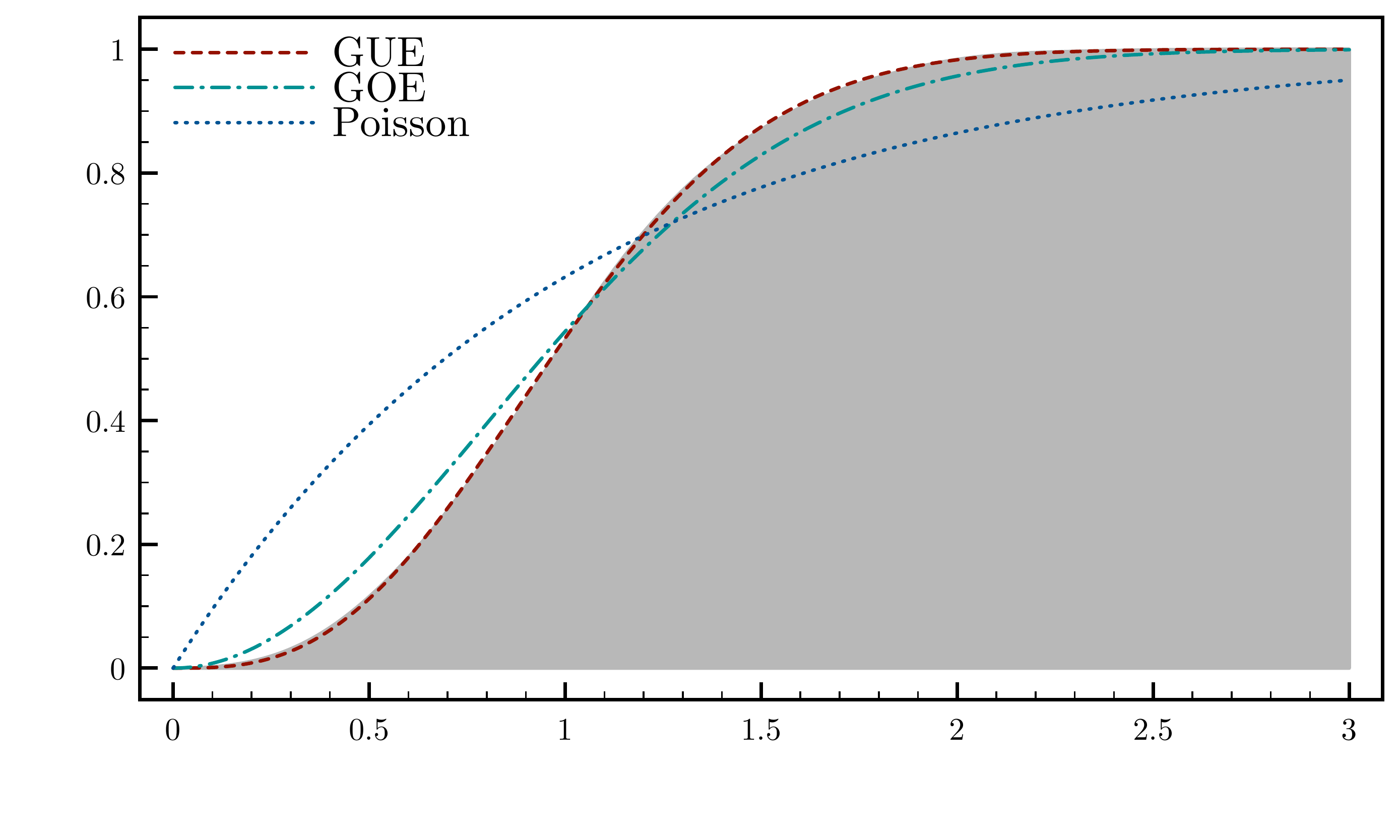}\hspace{0.5cm}\includegraphics[width=0.48\textwidth]{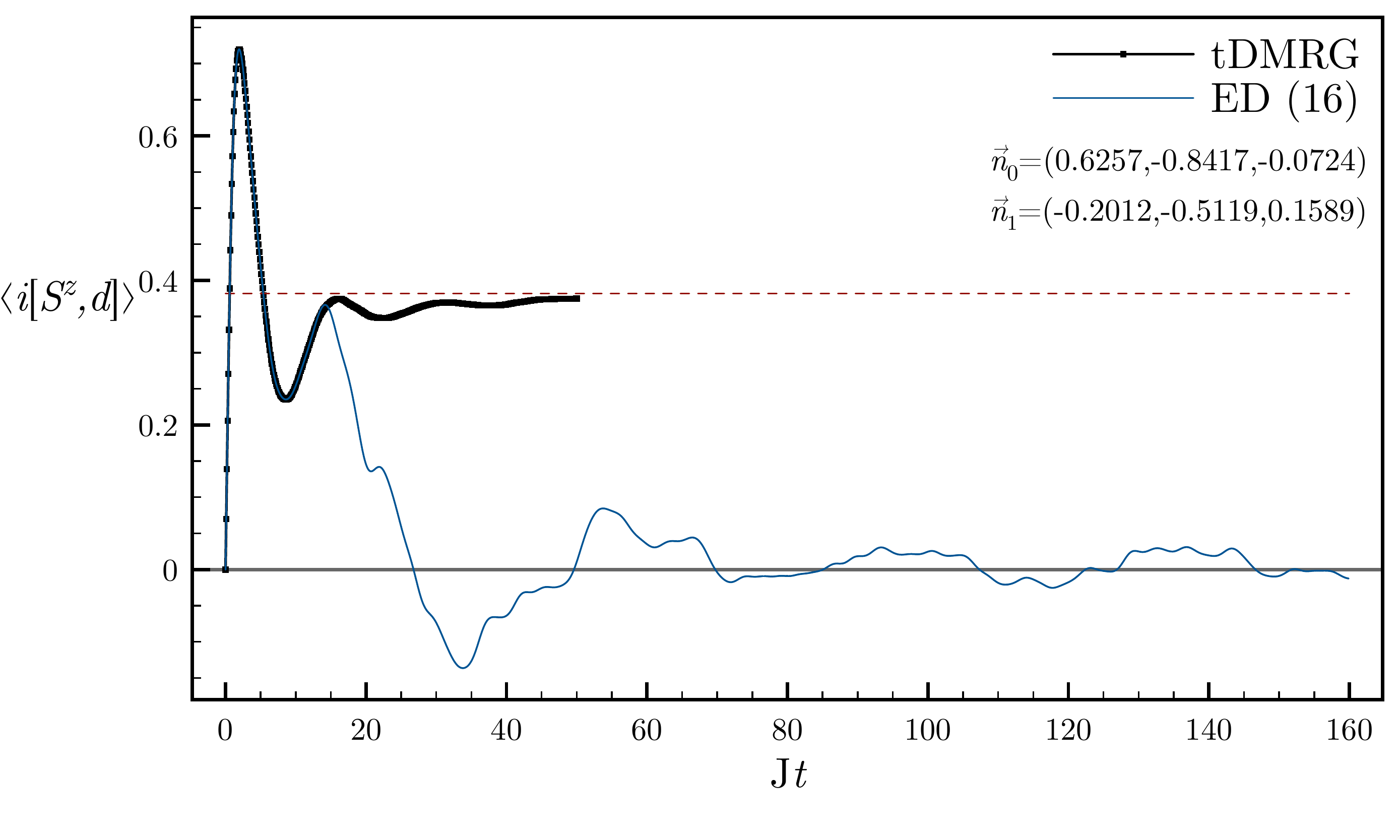}
\caption{The same as in \fref{fig:1} for the defect \eref{en:def3}.}\label{fig:3}
\end{center}
\end{figure}

We report the data for three defects:
\begin{enumerate}
\item \label{en:def1}$\vec n_0= (\sqrt{2}, 1, \frac{\pi}{4})$, $\vec n_1=(0,0,0)$;
\item \label{en:def2}$\vec n_0= (\frac{1}{2}, 0, \frac{\pi}{4})$, $\vec n_1=(0, \frac{1}{2}, \frac{\pi}{6})$;
\item \label{en:def3}$\vec n_0=(0.6257, -0.8417, -0.0724)$, $\vec n_1=(-0.2012, -0.5119, 0.158911)$.
\end{enumerate}
These are supposed to break integrability.
However,  the first defect preserves reflection symmetry about site $0$, and our estimate of the energy level spacing distribution is neither Gaussian (expected in generic models\footnote{For generic Hamiltonians, the cumulative nearest-neighbor spacing distribution (of the unfolded energy spectrum, with the mean normalized to unity) is believed~\cite{Dyson, BGS} to approach
\begin{itemize}
\item[-] a Gaussian unitary ensemble $P(\delta \epsilon<s)=\mathrm{erf}\bigl(\frac{2 s}{\sqrt{\pi }}\bigr)-\frac{4 s}{\pi }e^{-\frac{4 s^2}{\pi }}$ if the Hamiltonian is not invariant under time reversal;
\item[-] a Gaussian orthogonal ensemble $P(\delta \epsilon<s)=1 - e^{-\frac{\pi}{4} s^2}$ if the Hamiltonian is invariant under time reversal and the square of the time reversal operator is equal to $1$;
\item[-]  a Gaussian symplectic ensemble $P(\delta \epsilon<s)=\mathrm{erf}\bigl(\frac{8 s}{3 \sqrt{\pi }}\bigr)-\frac{16  s (128 s^2+27 \pi )}{81 \pi ^2}e^{-\frac{64 s^2}{9 \pi }}$ if the Hamiltonian is invariant under time reversal and the square of the time reversal operator is equal to $-1$.
\end{itemize}}) nor Poisson (expected in integrable models\footnote{For Hamiltonians of integrable models, the cumulative nearest-neighbor spacing distribution approaches a Poisson ensemble $P(\delta \epsilon<s)=1 - e^{-s}$~\cite{Berry}.}) - \fref{fig:1} (left). 
For the second defect - \fref{fig:2} (left) - we find a fair agreement with the Gaussian unitary ensemble. In order to reduce even more the risk of hidden symmetries, 
we generated the coefficients of the third defect randomly; the level spacing is now perfectly described by a Gaussian unitary ensemble  - \fref{fig:3} (left).  

Figures~\ref{fig:1},\ref{fig:2}, and \ref{fig:3} (right) show the time evolution of the local observable $i[S^z,d]$ in the three cases. The tDMRG data~\cite{Mario} leave almost no doubt that the anomalous defect current is nonzero. On the other hand, as expected, at times larger than the time necessary to traverse the chain, the observable starts oscillating around (and, roughly, also approaching) zero. 

We stress that the support of the observable ($i[S^z,d]$) coincides with the support of the defect, and the norm of the defect is $O(1)$; apparently, in the thermodynamic limit there is no small energy scale which might justify the approximate relaxation to a prethermalization plateau~\cite{preT}. 
Thus, having assumed \eref{eq:localT}, we conclude that, in all the cases considered, local observables do \emph{not} thermalize. 

\subsection{LQSS approaching the defect}%

We conclude with a final remark on the construction of the LQSS in the presence of defects.
For defects preserving integrability, Ref.~\cite{BF:16}  conjectured
\be\label{eq:boh}
\fl \qquad\qquad \lim_{\zeta\rightarrow 0^\pm}\lim_{t\rightarrow\infty\atop \ell =\zeta t}\braket{\Psi_0|e^{i H^{[1]} t} O_\ell  e^{-i H^{[1]} t}|\Psi_0}=\lim_{\ell\rightarrow\pm \infty}\lim_{t\rightarrow \infty}\braket{\Psi_0|e^{i H^{[1]} t} O_\ell  e^{-i H^{[1]} t}|\Psi_0}
\ee
and
\be\label{eq:NESSform}
\lim_{t\rightarrow\infty}\braket{\Psi_0|e^{i H^{[1]} t} O_\ell  e^{-i H^{[1]} t}|\Psi_0}=\frac{1}{Z}\tr[e^{\sum_n\lambda_n^{\rm NESS} Q^{[1](n)}}O_\ell]\, .
\ee
Here $Q^{[1](n)}$ are the (quasi-)local conservation laws of $H^{[1]}$ in the infinite line, which include both deformed and hybridized charges;  $\lambda_n^{\mathrm{NESS}}$ are real parameters.
Imposing the form \eref{eq:NESSform} is an alternative to imposing the boundary conditions corresponding to the LQSS in the limit $\zeta\rightarrow 0^\pm$; 
the values of $\lambda_n^{\mathrm{NESS}}$ turn out to be fixed by the continuity equations inside the light-cone and the boundary conditions outside. 

Both \eref{eq:boh} and \eref{eq:NESSform} have proven to be correct in the quantum Ising model with a defect that cuts the interactions between two sites~\cite{BF:16}. 
On the other hand, for generic defects, the results of the previous subsection cast doubt even on \eref{eq:NESSform}, which was a rather expected condition. Concerning \eref{eq:boh}, we note that weaker conditions could be sufficient to determine the LQSS. 
For example, a direct consequence of the continuity equation for a charge $Q^{[0]}$ of the unperturbed model is
\be\label{eq:limrt}
\lim_{r\rightarrow\infty}\lim_{t\rightarrow\infty}\tr [\rho_t (J^{[0]}_{-r}[Q^{[0]}]-J^{[0]}_r[Q])]=\mathcal A_d[{Q^{[0]}},\rho_0]\, .
\ee
If conjecture \eref{eq:boh} holds just for the currents, we find an equation for the current discontinuity $\Delta J^{[0]}[Q^{[0]}]$
\be\label{eq:limz0}
\fl\qquad\qquad \Delta J^{[0]}[Q^{[0]}]\overset{def}{=} \Bigl[\lim_{\zeta\rightarrow 0^+}-\lim_{\zeta\rightarrow 0^-}\Bigr]\lim_{t\rightarrow\infty\atop \ell =\zeta t}\braket{\Psi_t|J^{[0]}_{\ell}[Q^{[0]}]|\Psi_t}=-\mathcal A_d[{Q^{[0]}},\rho_0]\, .
\ee
Since deformed charges do not have anomalous defect currents, the currents associated with deformed charges develop removable discontinuities (as a function of $\zeta$) across the defect $\zeta\approx 0$. This conditions could be used to fix, or, at least, to constrain,  the parameters in \eref{eq:NESSform}. 

We leave to future works the assessment of the applicability of \eref{eq:NESSform} and \eref{eq:limz0} in the presence of generic defects.

\section{Conclusions}\label{s:conc}  %

We reviewed the structure of the local conservation laws in integrable models and computed the charge currents in generic noninteracting \mbox{spin-$\frac{1}{2}$} chains. We discussed the role of conservation laws and currents in the description of the late-time dynamics after quantum quenches, focusing in particular on the effects of a localized defect. We showed that the late-time expectation values of local observables can not be always described by a stationary state, and pointed out the failure of the diagonal ensemble, both in the presence of defects and when the initial state is not homogeneous.  We analyzed defects which are supposed to break integrability, obtaining strong indications that the stationary behavior of observables is not thermal. This leaves several open questions:
\begin{itemize}
\item We obtained a simple formal expression for the currents in the noninteracting case; what about interacting integrable models?
\item Are the localized defects considered in \sref{ss:genloc} sufficiently generic to break integrability?
\item Even if integrability is broken, 
are there quasi-local charges surviving the defect? Are there charges quasi-localized around the position of the defect?
\item Invoking the absence of long-range order in one-dimensional systems, we have assumed \eref{eq:localT}; could there be exceptions?
\item Which boundary conditions must be imposed around a generic defect in order to obtain the locally-quasi-stationary-state describing the late-time dynamics inside the light-cone?
\end{itemize}

\ack{
I am indebted to Mario Collura for having shown me the results of many tDMRG simulations in interacting models with localized defects, as well as for generously giving me the tDMRG data reported in Figs \ref{fig:1}, \ref{fig:2}, and \ref{fig:3}.  
I thank Bruno Bertini, Andrea De Luca, Jacopo De Nardis, Fabian Essler, and Giuseppe Mussardo for insightful discussions. Finally, I would like to give special thanks to Pasquale Calabrese for having encouraged the writing of this article.
This work was supported by LabEX ENS-ICFP:ANR-10-LABX-0010/ANR-10-IDEX-0001-02 PSL*.}

\appendix

\section[\hspace{2cm}]{Noninteracting spin-$\frac{1}{2}$ chains: further details}  %
\label{a:nonint}                                                                                                        %

In this appendix, we collect some details that could be useful especially to the reader interested in exceptional situations in which the set of the (quasi-)local conservation laws is not abelian. 

\subsubsection*{On the representations of the symbol.}  %

The $(y\kappa)$-symbol of a block  circulant matrix with $(2\kappa)$-by-$(2\kappa)$ blocks $\hat c_{\kappa}^j$ is a $y$-by-$y$ block matrix with the blocks (\emph{cf.} \eref{eq:symb})
\be
\fl\qquad[\hat c_{y \kappa}(z)]_{i' j'}=\hat c_{\kappa}^{j'-i'}+\sum_{n=1} (z^n\hat c_{\kappa}^{\,n y +j'-i'} +z^{-n}\hat c_{\kappa}^{\,-n y+j' -i'} )\qquad i',j'=1,\dots, y\, .
\ee
The block-elements depend only on the difference of the indices and satisfy
\be
[\hat c_{y \kappa}(z)]_{j+1,1}=z [\hat c_{y \kappa}(z)]_{j y}\, .
\ee
Using the terminology of \cite{factCirc}, $\hat c_{y \kappa}(z)$ is a $z$-factor block circulant  matrix of type $(y,2\kappa)$.
The spectrum $\sigma[\hat c_{y \kappa}(z)]$ of $\hat c_{y \kappa}(z)$ is easily obtained, and is given by
\be
\sigma[\hat c_{y \kappa}(z)]=\bigcup_{j=0}^{y-1}\sigma[\hat c_{\kappa}(e^{\frac{2\pi j}{y}} z^{\frac{1}{y}})]\, .
\ee
For example, the one-site XY+DM symbol \eref{eq:XYsymb} has the eigenvalues
\be
\fl\qquad\qquad \sigma[\hat h_{1}^{(\gamma,h,D)}(z)]=\Bigl\{{\mathrm J} D\frac{z-z^{-1}}{2i} \pm  \sqrt{\Bigl(h-\frac{z+z^{-1}}{2}\Bigr)^2-\gamma^2\Bigl(\frac{z-z^{-1}}{4}\Bigr)^2}\Bigr\}\, .
\ee
Thus, the eigenvalues of the two-site XY+DM symbol are
\begin{eqnarray}\label{eq:deg2}
\fl \sigma[\hat h^{(\gamma,h,D)}_{2}(z)]=\Bigl\{
&{\mathrm J} D\frac{z^{1/2}-z^{-1/2}}{2i} \pm  \sqrt{\Bigl(h-\frac{z^{1/2}+z^{-1/2}}{2}\Bigr)^2-\gamma^2\Bigl(\frac{z^{1/2}-z^{-1/2}}{4}\Bigr)^2},\nonumber\\
&{\mathrm J} D\frac{z^{-1/2}-z^{1/2}}{2i} \pm  \sqrt{\Bigl(h+\frac{z^{1/2}+z^{-1/2}}{2}\Bigr)^2-\gamma^2\Bigl(\frac{z^{1/2}-z^{-1/2}}{4}\Bigr)^2}
\Bigr\}\, .
\end{eqnarray}
Remarkably, if $h=0$ and  $D=0$ (XY Hamiltonian in zero field), for generic $z$ the spectrum is doubly degenerate.

\subsection*{Local conservation laws} %

Let $Q$ be a generic noninteracting charge and both $Q$ and the Hamiltonian $H$ be $\kappa$-site shift invariant. By rewriting  $Q$ and $H$ as in \eref{eq:spinM}, we find
\be
\fl \quad  0=[H,Q]=\frac{1-\Pi^z}{2}\frac{1}{4}\sum_{\ell,n=1}^{2L} a_\ell [\mathcal H^+,\mathcal Q^+]_{\ell n} a_n+\frac{1+\Pi^z}{2}\frac{1}{4}\sum_{\ell,n=1}^{2L} a_\ell [\mathcal H^-,\mathcal Q^-]_{\ell n} a_n\, ,
\ee
where we used $\{a_\ell, a_n\}=2\delta_{\ell n}$.
Since $\mathcal Q^\pm$ and $\mathcal H^\pm$ are \mbox{block-(anti-)circulant} matrices, their product is \mbox{block-(anti-)circulant} with $\kappa$-site symbol given by the product of the symbols. In conclusion we find~\cite{RDM, F:super}
\be\label{eq:charge}
[H,Q]=0\Longleftrightarrow [\hat h_{\kappa}(z),\hat q_\kappa(z)]=0\, .
\ee
Given the symbol $\hat q_\kappa(z)$, the corresponding charge follows directly from \eref{eq:symb}  and \eref{eq:spinM}.
\subsubsection*{Abelian case.}
In standard cases, for generic $z$, the symbol $\hat h_\kappa(z)$ of the Hamiltonian  is nondegenerate. 
Then, from \eref{eq:charge} it follows that a smooth symbol $\hat q_\kappa(z)$ must be a function of $\hat h_\kappa(z)$. Since $\hat h_\kappa(z)$ is a $(2\kappa)$-by-$(2\kappa)$ matrix, $\hat q_\kappa(z)$ can be recast as a polynomial of $\hat h_\kappa(z)$
\be\label{eq:abelian}
\hat q_\kappa(z)=\sum_{j=0}^{2\kappa-1} \alpha_{\kappa, j}(z)i^{j+1} [\hat h_\kappa(z)]^j\, .
\ee
By virtue of \eref{eq:symbprop}, the coefficients $\alpha_{\kappa, j}(z)$ are real and satisfy
\be
\alpha_{\kappa j}(1/z)=(-1)^{j-1}\alpha_{\kappa, j}(z)\, .
\ee
The conservation law is \emph{local} if $\hat q_\kappa(z)$ is also a polynomial in $z$ and $1/z$; since the Hamiltonian is local, namely $\hat h_\kappa(z)$ is already a polynomial, locality implies that also the coefficients $\alpha_{\kappa, j}(z)$ are polynomials in $z$ and $1/z$.

\subsubsection*{Non-abelian case.} %

If the $\kappa$-site symbol of the Hamiltonian is degenerate for generic $z$, there are other charges besides \eref{eq:abelian}: the powers of $\hat h_\kappa(z)$ can not resolve the degeneracy of the spectrum of $\hat h_\kappa(z)$. 

For $\kappa=1$, this is only possible if $\hat h_{1}(z)\propto \mathrm I_2$, \emph{e.g.} for the DM interaction~\eref{eq:DM}, which has $\hat h^{(0,0,D)}_{1}(z)=D\imath^{(0,-)}_{1}(e^{i k})$. As pointed out in \sref{s:charges},  the XY Hamiltonian $H^{\rm XY}_{\gamma,h}$ commutes with $H_D^{\rm DM}$  for any value of the parameters. However, for $\gamma\neq 0$ and $h\neq h'$, $[H^{\rm XY}_{\gamma,h},H^{\rm XY}_{\gamma,h'}]\neq 0$ and, in turn, the corresponding symbols do non commute with one another.  Consequently, the set of the local conservation laws of the DM interaction is non-abelian.  
A sensible choice of independent charges is generated by the following symbols
\be
\sin(n k)\mathrm I,\,\sin(n k)\sigma^x,\,\cos(n k)\sigma^y,\,\sin(n k)\sigma^z\, ,
\ee
where $n$ is integer.

For $\kappa=2$, there are more interesting cases. For example, we have shown that for $h=0$ the symbol of the XY Hamiltonian \eref{eq:XY} is doubly degenerate (\emph{cf}. \eref{eq:deg2}) and, in particular, $\hat h^{(\gamma,0,0)}_{1}(-e^{i k})=-\hat h_{1}^{(\gamma,0,0)}(e^{i k})$. 
The symbols resolving the degeneracy generate charges that do not commute with one another. Being the degeneracy independent of $z$ (\emph{cf}. \eref{eq:deg2}), such charges can be chosen to be local, as originally shown in Ref.~\cite{F:super}.
Moreover, since the new charges do not have a one-site symbol, they break one-site shift invariance. 
They have been classified according to their transformation rules under chain inversion $\rm R$ and spin flip $\Pi^x: O\mapsto \prod_\ell\sigma_\ell^x O \prod_\ell\sigma_\ell^x$, which act on the symbols as follows:
\begin{eqnarray}
{\rm R}&:\quad \hat o_\kappa(e^{i k})\mapsto [\Sigma_\kappa^x\otimes \sigma^y]\hat o_\kappa(e^{-i k})[\Sigma_\kappa^x\otimes \sigma^y]\nn
\Pi^x&:\quad \hat o_\kappa(e^{i k})\mapsto[\Sigma_\kappa^z\otimes \sigma^z]\hat o_\kappa((-1)^\kappa e^{i k})[\Sigma_\kappa^z\otimes \sigma^z]
\end{eqnarray}
where
\be
[\Sigma_\kappa^x]_{i j}=\delta_{i+j,1+\kappa}\qquad [\Sigma_\kappa^z]_{i j}=(-1)^{\kappa+1-i}\delta_{i j}\qquad 1\leq i,j\leq \kappa\, .
\ee
 The symbols of the local charges are then given by:
\begin{eqnarray}\label{eqa:QXY}
\fl\qquad\qquad  \hat i_2^{(n,++)}(e^{i k})=\cos(n k)[\sigma^xe^{i\frac{k}{2}\sigma^z}]\otimes \hat h_{1}^{(\gamma,0,0)}(e^{i k/2})&\sim \quad \hat i_1^{(2n,+)}(e^{i k})\nn
\fl\qquad\qquad \hat i_2^{(n,+-)}(e^{i k})=\cos((n+1/2) k)\mathrm I\otimes \hat h_{1}^{(\gamma,0,0)}(e^{i k/2})&\sim\quad  \hat i_1^{(2n+1,+)}(e^{i k})\nn
\fl\qquad\qquad \hat i_2^{(n,-+)}(e^{i k})={\mathrm J}\sin((n+1) k)\mathrm I\otimes \mathrm I&\sim\quad  \hat i_1^{(2n+1,-)}(e^{i k})\nn
\fl\qquad\qquad \hat i_2^{(n,--)}(e^{i k})={\mathrm J}\sin((n+1/2) k)[\sigma^xe^{i\frac{k}{2}\sigma^z}]\otimes\mathrm I&\sim\quad \hat i_1^{(2n,-)}(e^{i k})\nn
\fl\qquad\qquad \hat y_2^{(n,++)}(e^{i k})=\cos(n k)[\sigma^ye^{i\frac{k}{2}\sigma^z}]\otimes [i\sigma^z\hat h_{1}^{(\gamma,0,0)}(e^{i k/2})]\nn
\fl\qquad\qquad \hat y_2^{(n,+-)}(e^{i k})={\mathrm J}\cos((n+1/2) k)[\sigma^ye^{i\frac{k}{2}\sigma^z}]\otimes \sigma^z\nn
\fl\qquad\qquad \hat y_2^{(n,-+)}(e^{i k})={\mathrm J}\sin((n+1) k)\sigma^z \otimes \sigma^z\nn
\fl\qquad\qquad \hat y_2^{(n,--)}(e^{i k})=\sin((n+1/2) k)\sigma^z  \otimes [i\sigma^z\hat h_{1}^{(\gamma,0,0)}(e^{i k/2})]
\end{eqnarray}
for generic integer $n\geq 0$. Here we used $\{\sigma^z,\hat h_{1}^{(\gamma,0,0)}(e^{i k/2})\}=0$ (\emph{cf.} \eref{eq:XYsymb} with $h=D=0$). 
The charges $I^{(2n+\frac{1-s_1s_2}{2},s1)}$ and $Y^{(n,s_1,s_2)}$, corresponding to $\hat i_2^{(n,s_1s_2)}(e^{i k})$ and $\hat y_2^{(n,s_1s_2)}(e^{i k})$ respectively,  take sign $s_1$ under $\rm R$ and sign $s_2$ under $\Pi^x$. 
The former class is one-site shift invariant, and, on the right hand side of \eref{eq:QXY}, we also reported the corresponding one-site symbols. The $Y$ charges are instead odd under a shift by one site. 
This can be verified using that the one-site shift operator $\rm U$ acts on a two-site symbol as follows
\be
{\rm U}:\quad \hat o_2(z)\mapsto [(\sigma^xe^{i\frac{k}{2}\sigma^z})\otimes\mathrm I]\, \hat o_2(z) [(\sigma^xe^{i\frac{k}{2}\sigma^z})\otimes\mathrm I]\, .
\ee 

\paragraph{Loop algebra.}   %

As also shown in Ref.~\cite{Bruno}, one can reorganize the set of the local conservation laws into non-hermitian charges $T^{(n,\alpha)}$ with the following symbols 
\begin{eqnarray}\label{eqa:loop}
\hat t_2^{(n,0)}(e^{i k})=\hat i_2^{++}(e^{i k})e^{-i n k\, \hat i_2^{++}(e^{i k})}\nn
\hat t_2^{(n,1)}(e^{ik})=\hat y_2^{+-}(e^{i k}) e^{i\frac{|k|}{2}\hat y_2^{-+}(e^{i k})}e^{-i n k\, \hat i_2^{++}(e^{i k})}\nn
\hat t_2^{(n,2)}(e^{ik})=\hat y_2^{++}(e^{i k})e^{-i n k\, \hat i_2^{++}(e^{i k})}\nn
\hat t_2^{(n,3)}(e^{ik})=\hat i_2^{+-}(e^{i k})  e^{i\frac{|k|}{2}\hat y_2^{-+}(e^{i k})}e^{-i n k\, \hat i_2^{++}(e^{i k})}\, ,
\end{eqnarray}
where $\hat i_2^{s s'}(e^{i k})=\mathrm{sgn}[\hat i_2^{(0,ss')}(e^{i k})]$ and $\hat y_2^{s s'}(e^{i k})=\mathrm{sgn}[\hat y_2^{(0,s s')}(e^{i k})]$.
The operators $T^{(n,\alpha)}=[T^{(-n,\alpha)}]^\dag$ are quasi-local and generate the loop algebra
\be\label{eqa:loop}
{}[T^{(m,\alpha)},T^{(n,\beta)} ]=2i \sum_\gamma\epsilon_{\alpha\beta\gamma}  T^{(m+n,\gamma)}\, ,
\ee
where
\be
\epsilon_{\alpha\beta\gamma}=(1-\delta_{\alpha 0})(1-\delta_{\beta 0})(1-\delta_{\gamma 0})\varepsilon_{\alpha\beta\gamma}\, ,
\ee
and $\varepsilon_{\alpha\beta\gamma}$ is the Levi-Civita symbol. 
The symbols \eref{eqa:loop} also satisfy
\be
\{\hat t^{(m,i)}(e^{i k}),\hat t^{(n,j)}(e^{i k}) \}=2\delta_{i j}e^{-i (n+m) k\, \hat i_2^{++}(e^{i k})}\qquad i,j=1,2,3
\ee

\subsection*{Expectation values in a macro-state: non-abelian case}\label{a:nonab}%

If the set of charges is non-abelian, the parametrization in terms of  densities is more complicated. Let us consider for example the XY model in zero field ($h=0$), which we have shown to have a non-abelian set of local conservation laws. 
Using the non-hermitian representation \eref{eq:loop} for the set of charges, the most general two-site symbol of the correlation matrix in a stationary state can be written as
\be
\hat\Gamma_2(z)=\sum_{\alpha=0}^3 \int\frac{\mathrm d p}{2\pi} v_\alpha(p)\hat t^{(\alpha)}(z,e^{ip})
\ee
where
\be
\hat t^{(\alpha)}(z,w)=\sum_n w^n\hat t^{(n,\alpha)}(z)=\hat t^{(0,\alpha)}(z)\hat t(z,w)
\ee
\be
\hat t(e^{i k},e^{i p})=\frac{\mathrm I+ \hat i_2^{++}(e^{i k})}{2}2\pi\delta(p-k)+\frac{\mathrm I- \hat i_2^{++}(e^{i k})}{2}2\pi\delta(p+k)\, .
\ee
The functions $v_\alpha(k)$ are real and satisfy
\be\label{eq:cond0}
-1\leq |v_0(k)|\pm \sqrt{\sum_{i=1}^3 [v_i(k)]^2}\leq 1
\ee
This follows from the following facts:
\begin{itemize}
\item the eigenvalues of the symbol of a correlation matrix lie in the interval $[-1,1]$;
\item $\hat t(e^{i k},e^{i p})$ commutes with all the other symbols and has eigenvalues $2\pi \delta(k\pm p)$;
\item $\{\hat t^{(j)}(z,e^{i p}),\hat t^{(k)}(z,e^{i \tilde p})\}=4\pi \delta(p-\tilde p)\delta_{j k}\hat t(z,e^{i p})$, for $j,k\in\{1,2,3\}$. 
\end{itemize}
Isolating the zero term of the sum in $\hat\Gamma_2(z)$ and taking the square of the remainder result in \eref{eq:cond0}. 

A possible parametrization compatible with \eref{eq:root} consists of two densities and an auxiliary normalized vector function $\vec u(k)$ that selects the particular abelian subset of charges the correlation matrix belongs to:
\begin{eqnarray}
v_0(k)=2\pi[\rho_{1,p}(k| u)+\rho_{2,p}(k| u)]-1\nn
v_i(k)=2\pi[\rho_{1,p}(k| u)-\rho_{2,p}(k| u)] u_i(k)\nn
\sum\nolimits_{j=1}^3 u_j^2(k)=1\, .
\end{eqnarray}
That is to say
\be\label{eq:Gamma2t}
\fl\qquad  \hat\Gamma_2(z)=\sum_{j=1}^2\int\mathrm d p\rho_{j,p}(k|  u)\Bigl[\hat t^{(0)}(z,e^{ip})+(3-2j)\hat t_{u}(z,e^{ip})]-\int\frac{\mathrm d p}{2\pi}\hat t^{(0)}(z,e^{ip})\, ,
\ee
where 
$
\hat t_{u}(e^{i k},e^{ip})=\sum_{j=1}^3 u_j(p) \hat t^{(j)}(e^{i k},e^{ip})\, .
$

\section[\hspace{2cm}]{Currents}\label{a:current}  %

In this appendix, we prove the formal expression reported in \sref{s:current} for the currents associated with the (noninteracting) local conservation laws.  

Let us consider a chain with an even number of sites. 
We write the matrix $\mathcal H$ associated with the Hamiltonian in block diagonal form, the blocks having half of the total size. Using translational invariance and Hermiticity, we have
\be
\mathcal H^{\pm}=\left(\begin{array}{cc}\mathcal H_{1/2}^{\rm{OBC}}&\mathcal W_H^\pm\\
\pm \mathcal W_H^\pm&\mathcal H_{1/2}^{\rm{OBC}}
\end{array}\right)\, ,
\ee
where $\mathcal H_{1/2}^{\rm{OBC}}$ is the matrix corresponding to the Hamiltonian if the chain is halved and open boundary conditions are imposed $H^{\rm OBC}_{1/2}=\frac{1}{4}\sum_{\ell, n}a_\ell \mathcal H_{1/2}^{\rm{OBC}} a_n$. 
We do the same for a generic local conservation law $Q$:
\be
\mathcal Q^{\pm}=\left(\begin{array}{cc}\mathcal Q_{1/2}^{\rm{OBC}}&\mathcal W_Q^\pm \\
\pm \mathcal W_Q^\pm &\mathcal Q_{1/2}^{\rm{OBC}}
\end{array}\right)\, .
\ee
Since this is associated with a charge, $[\mathcal H^{\pm},\mathcal Q^{\pm}]=0$, and the following identities hold:
\begin{eqnarray}\label{eq:cond}
[\mathcal H_{1/2}^{\rm{OBC}},\mathcal Q_{1/2}^{\rm{OBC}}]=\pm[\mathcal W_Q^\pm,\mathcal W_H^\pm]\nonumber \\
{}[\mathcal H_{1/2}^{\rm{OBC}},\mathcal W_Q^\pm]=[\mathcal Q_{1/2}^{\rm{OBC}},\mathcal W_H^\pm]
\end{eqnarray}
If we indicate the charge density by $Q_\ell$, the charge restricted to half chain can be represented as follows
\be
\fl \qquad \sum_{\ell=1}^{L/2} Q_\ell=\frac{1-\Pi^z}{8}\vec a^\dag \left(\begin{array}{cc}\mathcal Q_{1/2}^{\rm{OBC}}&\frac{1}{2}\mathcal W_Q^+\\
\frac{1}{2}\mathcal W_Q^+&0
\end{array}\right) \vec a+\frac{1+\Pi^z}{8}\vec a^\dag \left(\begin{array}{cc}\mathcal Q_{1/2}^{\rm{OBC}}&\frac{1}{2}\mathcal W_Q^-\\
-\frac{1}{2}\mathcal W_Q^-&0
\end{array}\right) \vec a\, ,
\ee
where we introduced the vector notations $[\vec a]_\ell =a_\ell$.
The current density $J_\ell[Q]$ satisfies the continuity equation
\be
J_{L/2+1}[Q]-J_{1}[Q]=-i\Bigl[H,\sum_{\ell=1}^{L/2} Q_\ell\Bigr]\, .
\ee
Expanding $J_\ell[Q]$ as in  \eref{eq:spinM} gives 
\be
\mathcal J^\pm_{L/2+1}[Q]-\mathcal J^\pm_{1}[Q]= -i\left[\left(\begin{array}{cc}\mathcal H_{1/2}^{\rm{OBC}}&\mathcal W_H^\pm\\
\pm \mathcal W_H^\pm&\mathcal H_{1/2}^{\rm{OBC}}
\end{array}\right),\left(\begin{array}{cc}\mathcal Q_{1/2}^{\rm{OBC}}&\frac{1}{2}\mathcal W_Q^\pm\\
\pm \frac{1}{2}\mathcal W_Q^\pm&0
\end{array}\right)\right]\, ,
\ee
where $\mathcal J_\ell^\pm[Q]$ are the matrices associated with $J_{\ell}[Q]$.
From \eref{eq:cond} it follows
\be\label{eq:commant}
\mathcal J_{L/2+1}^\pm[Q]-\mathcal J_{1}^\pm[Q]=\frac{i}{2}\left(\begin{array}{cc}
-[\mathcal H_{1/2}^{\rm{OBC}},\mathcal Q_{1/2}^{\rm{OBC}}]&\{\mathcal W_H^\pm,\mathcal Q_{1/2}^{\rm{OBC}}\}\\
\mp\{\mathcal W_H^\pm,\mathcal Q_{1/2}^{\rm{OBC}}\}&[\mathcal H_{1/2}^{\rm{OBC}},\mathcal Q_{1/2}^{\rm{OBC}}]
\end{array}\right)\, .
\ee
Let $\kappa$ be large enough so that $\mathcal H^\pm$ can be seen as a block-tridiagonal (anti-)circulant matrix with $(2\kappa)$-by-$(2\kappa)$ blocks. For the XY Hamiltonian with the DM interaction one can choose $\kappa=1$; for Hamiltonians with longer range interactions it could be necessary to choose larger values of $\kappa$. 
We have
\begin{eqnarray}
[\mathcal W_H^\pm]_{\ell n}&=\delta_{\ell, \frac{L}{2\kappa}-1 }\delta_{n, 0}\hat h^+_\kappa\pm \delta_{\ell, 0}\delta_{n, \frac{L}{2\kappa}-1 }\hat h^-_\kappa\nonumber\\
{}[\mathcal H_{1/2}^{\rm{OBC}}]_{\ell, n}&=\delta_{\ell, n-1}\hat h^-_\kappa+\delta_{\ell, n}\hat h^0+\delta_{\ell, n+1}\hat h^+_\kappa\nonumber\\
{}[\mathcal Q_{1/2}^{\rm{OBC}}]_{\ell, n}&=\hat q_\kappa^{n-\ell}\qquad \qquad \qquad\qquad \qquad\qquad 0\leq \ell, n<\frac{L}{2\kappa}\, .
\end{eqnarray}
The commutators and anticommutators in \eref{eq:commant} can be written in a rather simple form by exploiting $[\mathcal H_{1/2}^{\pm},\mathcal Q_{1/2}^{\pm}]=0$, where $\mathcal O_{1/2}^\pm$ are the matrices corresponding to a given shift invariant operator $O$ for the chain halved. We find
\begin{eqnarray}
\fl {}[\mathcal H_{1/2}^{\rm{OBC}},\mathcal Q_{1/2}^{\rm{OBC}}]&=\left(\begin{array}{ccccc}
-\hat h^-_\kappa \hat q_\kappa^1+\hat q_\kappa^{-1}\hat h^+_\kappa&-\hat h^-_\kappa \hat q_\kappa^2&
\rightarrow &0&0\\
\hat q_\kappa^{-2}\hat h^+_\kappa&0&\cdots&0&0\\
\downarrow&\vdots&&\vdots&
\uparrow\\
0&0&\cdots&0&\hat q_\kappa^2\hat h^-_\kappa \\
0&0&\leftarrow
&-\hat h^+_\kappa\hat q_\kappa^{-2}&\hat q_\kappa^1\hat h^-_\kappa -\hat h^+_\kappa\hat q_\kappa^{-1}
\end{array}\right)\\
\fl {}\{\mathcal W_H^\pm ,\mathcal Q_{1/2}^{\rm{OBC}}\}&=\left(\begin{array}{ccccc}
0&0&\leftarrow
&\pm \hat h^-_\kappa \hat q_\kappa^{-1}&\pm (\hat h^-_\kappa \hat q_\kappa^0+ \hat q_\kappa^0\hat h^-_\kappa) \\
0&0&\cdots&0&\pm \hat q_\kappa^{-1}\hat h^-_\kappa \\
\uparrow
&\vdots&&\vdots&
\downarrow
\\
\hat q_\kappa^1\hat h^+_\kappa&0&\cdots&0&0\\
\hat h^+_\kappa\hat q_\kappa^0+\hat q_\kappa^0\hat h^+_\kappa&\hat h^+_\kappa\hat q_\kappa^1&
\rightarrow&0&0
\end{array}\right)\, ,
\end{eqnarray}
where the arrows indicate that the subsequent elements in the corresponding direction have the same form but the index $j$ of $\hat q_\kappa^j$ increases moving upwards and rightwards and decreases moving downwards and leftwards.
Plugging these expressions into \eref{eq:commant} gives
\begin{eqnarray}
\fl \mathcal J_{L/2+1}^\pm [Q]=\nonumber\\
\fl\qquad\quad \frac{i}{2}\left(\begin{array}{cccccccc}
0&\cdots&0&0&0&0&\cdots&0\\
\vdots&&\vdots&\uparrow&\uparrow&\vdots&&\vdots\\
0&\cdots&0&-\hat q_\kappa^2\hat h^-_\kappa&\hat q_\kappa^1\hat h^+_\kappa&0&\cdots&0\\
0&\leftarrow&-\hat h^+_\kappa \hat q_\kappa^{-2}&-\hat q_\kappa^1 \hat h^-_\kappa+\hat h^+_\kappa\hat q_\kappa^{-1}&\hat h^+_\kappa\hat q_\kappa^0+\hat q_\kappa0\hat h^+_\kappa&\hat h^+_\kappa\hat q_\kappa^1&\rightarrow&0\\
0&\leftarrow&-\hat h^-_\kappa \hat q_\kappa^{-1}&-\hat q_\kappa^0 \hat h^-_\kappa-\hat h^-_\kappa\hat q_\kappa^{0}&-\hat h^-_\kappa\hat q_\kappa^1+\hat q_\kappa^{-1}\hat h^+_\kappa&-\hat h^-_\kappa\hat q_\kappa^2&\rightarrow&0\\
0&\cdots&0&-\hat q_\kappa^{-1}\hat h^-_\kappa&\hat q_\kappa^{-2}\hat h^+_\kappa&0&\cdots&0\\
\vdots&&\vdots&\downarrow&\downarrow&\vdots&&\vdots\\
0&\cdots&0&0&0&0&\cdots&0
\end{array}\right).
\end{eqnarray}
Here we used that $\mathcal J^\pm_{L/2+1}[Q]$ and $\mathcal J^\pm_{1}[Q]$ are  localized around the middle of the chain and around the first site, respectively.
This is the matrix associated with the current density. A shift by $j$ in the indices of $\mathcal J_{L/2+1}^\pm [Q]$ corresponds to a shift in the chain by $j$ ``macro-sites'', \emph{i.e.} $\kappa j$ sites, that is to say, $[\mathcal J^\pm_{L/2+1-j}[Q]]_{\ell, n}=[\mathcal J^\pm_{L/2+1}[Q]]_{\ell+\kappa j, n+\kappa j}$ \mbox{((anti-)periodicity} in the block-indices is understood). 
Summing the current density over all the macro-sites gives the current. By translational invariance, the matrices associated with the current are block \mbox{(anti-)circulant}. Their elements are nothing but the sum of the block-elements of $\mathcal J_{L/2+1}^\pm [Q]$ over the block-diagonals, \emph{i.e.}
\be
\hat \jmath_\kappa^n=\frac{i}{2}\Bigl(\{\hat q_\kappa^{n-1},\hat h^+_\kappa\}-\{\hat q_\kappa^{n+1},\hat h_\kappa^-\}\Bigr)\, .
\ee
The associated symbol is (\emph{cf}. \eref{eq:symb})
\be
\fl\qquad \qquad\hat \jmath_\kappa(z)=\frac{i}{2}\sum_n z^n \bigl(\{\hat q_\kappa^{n-1},\hat h_\kappa^+\}-\{\hat q_\kappa^{n+1},\hat h_\kappa^-\}\bigr)=\frac{i}{2}\{\hat q(z),z\hat h^+-z^{-1}\hat h^-\}\, .
\ee
Using $iz\partial_z\hat h_\kappa(z)=i z\hat h_\kappa^+-i z^{-1}\hat h_\kappa^-$, one finally obtains \eref{eq:current}.

\section*{References} %

\end{document}